\documentclass[journal,draftcls,onecolumn,12pt,twoside]{IEEEtran}
\usepackage[noadjust]{cite}
\usepackage{graphicx,color}
\usepackage[dvipsnames]{xcolor}
\usepackage{amsmath,amsbsy,amsfonts,amssymb}
\usepackage{mathrsfs,bm}
\usepackage{mathtools}
\usepackage{subfigure}
\ifCLASSOPTIONcompsoc
\usepackage[caption=false,font=normalsize,labelfont=sf,textfont=sf]{subfig}
\else
\usepackage[caption=false]{subfig}
\fi

\IEEEoverridecommandlockouts

\newcommand{\rev}{\textcolor{black}}

\usepackage{acro}
\newtheorem{example}{Example}

\DeclareAcronym{AWGN}{short = AWGN ,long = additive white gaussian noise}
\DeclareAcronym{AoI}{short = AoI ,long = age of information}
\DeclareAcronym{PAoI}{short = PAoI ,long = peak age of information}
\DeclareAcronym{CDF}{short = CDF ,long = cumulative distribution function}
\DeclareAcronym{CRA}{short = CRA ,long = contention resolution ALOHA}
\DeclareAcronym{CRDSA}{short = CRDSA ,long = contention resolution diversity slotted ALOHA}
\DeclareAcronym{CP}{short = CP ,long = contention period}
\DeclareAcronym{CSA}{short = CSA ,long = coded slotted ALOHA}
\DeclareAcronym{C-RAN}{short = C-RAN ,long = cloud radio access network}
\DeclareAcronym{DAMA}{short = DAMA ,long = demand assigned multiple access}
\DeclareAcronym{DSA}{short = DSA ,long = diversity slotted ALOHA}
\DeclareAcronym{eMBB}{short = eMBB ,long = enhanced mobile broadband}
\DeclareAcronym{FEC}{short = FEC ,long = forward error correction}
\DeclareAcronym{GEO}{short = GEO ,long = geostationary orbit}
\DeclareAcronym{HAP}{short = HAP ,long = high-altitude platform,foreign-plural={}}
\DeclareAcronym{IC}{short = IC ,long = interference cancellation}
\DeclareAcronym{IoT}{short = IoT ,long = internet of things}
\DeclareAcronym{IRSA}{short = IRSA ,long = irregular repetition slotted ALOHA}
\DeclareAcronym{LEO}{short = LEO ,long = low Earth orbit}
\DeclareAcronym{M2M}{short = M2M ,long = machine-to-machine}
\DeclareAcronym{MAC}{short = MAC ,long = medium access control}
\DeclareAcronym{MPR}{short = MPR ,long = multi-packet reception}
\DeclareAcronym{MTC}{short = MTC ,long = machine-type communications}
\DeclareAcronym{mMTC}{short = mMTC ,long = massive machine-type communications}
\DeclareAcronym{NTN}{short = NTN ,long = non-terrestrial network,foreign-plural = {}}
\DeclareAcronym{PDF}{short = PDF ,long = probability density function}
\DeclareAcronym{PER}{short = PER ,long = packet error rate}
\DeclareAcronym{PLR}{short = PLR ,long = packet loss rate}
\DeclareAcronym{PMF}{short = PMF ,long = probability mass function}
\DeclareAcronym{RA}{short = RA ,long = random access}
\DeclareAcronym{RRH}{short = RRH ,long = remote radio head,foreign-plural = {}}
\DeclareAcronym{SA}{short = SA , long = slotted ALOHA}
\DeclareAcronym{SIC}{short = SIC ,long = successive interference cancellation}
\DeclareAcronym{SIR}{short = SIR ,long = signal to interference ratio}
\DeclareAcronym{SNIR}{short = SNIR ,long = signal-to-noise and interference ratio}
\DeclareAcronym{SINR}{short = SINR ,long = signal-to-interference and noise ratio}
\DeclareAcronym{SNR}{short = SNR ,long = signal-to-noise ratio}
\DeclareAcronym{TDM}{short = TDM ,long = time division multiplexing}

\newcommand{\pr}{\ensuremath{\mathbb P}}
\newcommand{\prob}{\ensuremath{\mathbb P}}
\newcommand{\expOp}{\ensuremath{\mathbb E}}

\newcommand{\tru}{\ensuremath{\mathsf S}}

\newcommand{\nodes}{\ensuremath{\mathsf U}}

\newcommand{\nodesRV}{\ensuremath{U}}
\newcommand{\nodesRVl}{\ensuremath{\nodesRV^{(\ell)}}}
\newcommand{\nodesRVll}{\ensuremath{\nodesRV^{(\ell+1)}}}
\newcommand{\nodesrv}{\ensuremath{u}}

\newcommand{\maxFrame}{\ensuremath{d_{\text{max}}}}
\newcommand{\cpRV}{\ensuremath{D}}
\newcommand{\cpRVl}{\ensuremath{\cpRV^{(\ell)}}}
\newcommand{\cpRVll}{\ensuremath{\cpRV^{(\ell+1)}}}
\newcommand{\cprv}{\ensuremath{d}}

\newcommand{\decRV}{\ensuremath{M}}
\newcommand{\decRVl}{\ensuremath{\decRV^{(\ell)}}}
\newcommand{\decrv}{\ensuremath{m}}

\newcommand{\pDGivenN}{\ensuremath{P_{\cpRV|\nodesRV}}}
\newcommand{\pMGivenN}{\ensuremath{P_{\decRV|\nodesRV}}}

\newcommand{\succRV}{\ensuremath{S}}
\newcommand{\succrv}{\ensuremath{s}}
\newcommand{\succRVl}{\ensuremath{\succRV^{(\ell)}}}

\newcommand{\pActGivenL}{{\ensuremath{\mathsf \gamma_{\cprv}}}}
\newcommand{\pActGen}{\ensuremath{\mathsf \gamma}}

\newcommand{\pAct}{\ensuremath{\gamma}}
\newcommand{\pTx}{\ensuremath{q}}
\newcommand{\pUpdate}{\ensuremath{\nu}}

\newcommand{\AoI}{\ensuremath{\Delta}}
\newcommand{\minAoI}{\ensuremath{\hat\Delta}}
\newcommand{\maxTru}{\ensuremath{\hat\tru}}

\newcommand{\pmc}[3]{\ensuremath{p_{#1}(#2,#3)}}

\newcommand{\unres}{\ensuremath{w}}
\newcommand{\collisions}{\ensuremath{c}}
\newcommand{\singletons}{\ensuremath{r}}
\newcommand{\au}{\ensuremath{i}}
\newcommand{\bu}{\ensuremath{{j}}}
\newcommand{\buall}{\ensuremath{{a}}}
\newcommand{\qu}{\ensuremath{h_\unres}}
\newcommand{\state}{\ensuremath{\mathrm{Pre}}}
\newcommand{\staten}[1]{\ensuremath{\state_{#1}}}
\newcommand{\stateafter}{\ensuremath{\mathrm{Pos}}}
\newcommand{\stateaftern}[1]{\ensuremath{\stateafter_{#1}}}
\newcommand{\probmn}[2]{\ensuremath{\beta(#1, #2)}}

\newcommand{\avgAoI}{\ensuremath{\bar{\AoI}}}
\newcommand{\initAoI}{\ensuremath{X}}
\newcommand{\initaoi}{\ensuremath{x}}
\newcommand{\interUpdate}{\ensuremath{Y}}
\newcommand{\ancMC}{\ensuremath{Z}}
\newcommand{\ancmc}{\ensuremath{z}}
\newcommand{\ancMCone}{\ensuremath{\ancMC^{(1)}}}
\newcommand{\ancMCl}{\ensuremath{\ancMC^{(\ell)}}}
\newcommand{\ancMCll}{\ensuremath{\ancMC^{(\ell+1)}}}
\newcommand{\succRVll}{\ensuremath{\succRV^{(\ell+1)}}}
\newcommand{\drift}{\ensuremath{\Xi}}

\newcommand{\decAtMaxFrame}{\ensuremath{\beta}}

\newcommand{\statDistCP}{\ensuremath{\pi_\cpRV}}
\newcommand{\statDistNodes}{\ensuremath{\pi_\nodesRV}}
\newcommand{\statDistDec}{\ensuremath{\pi_\decRV}}
\newcommand{\statDistAncMC}{\ensuremath{\pi_\ancMC}}

\begin{document}

\title{The Dynamic Behavior of Frameless ALOHA: \rev{Drift Analysis}, Throughput, and Age of Information}
\author{Andrea Munari \IEEEmembership{Senior Member, IEEE}, Francisco L\'azaro \IEEEmembership{Senior Member, IEEE}, Giuseppe Durisi \IEEEmembership{Senior Member, IEEE}, Gianluigi Liva \IEEEmembership{Senior Member, IEEE}
\vspace{-1.5em}
\thanks{A. Munari, F. L\'azaro, and G. Liva are with the Institute of Communications and Navigation, German Aerospace Center (DLR), 82234 Wessling, Germany (email: \{andrea.munari, francisco.lazaroblasco, gianluigi.liva\}@dlr.de)\\
G. Durisi is with the Department of Electrical Engineering, Chalmers University of Technology, Gothenburg, 41296, Sweden (e-mail: durisi@chalmers.se).
}
\thanks{This paper was presented in part at the 2021 Asilomar Conference on Signals, Systems, and Computers, Pacific Grove, CA,
USA~\cite{Munari21:Asilomar}.}
\thanks{The authors acknowledge the financial support provided  by the Swedish Research Council under grant 2021-04970, and by the Federal Ministry of Education and Research of Germany in the programme of "Souver\"an. Digital. Vernetzt." Joint project 6G-RIC, project identification number: 16KISK022.} 
} \maketitle
\thispagestyle{empty} \setcounter{page}{0}

\markboth
    {A. Munari et al.: The Dynamic Behavior of Frameless ALOHA: Drift Analysis, Throughput and AoI}
\maketitle

\begin{abstract}
We study the dynamic behavior of frameless ALOHA, both in terms of throughput
and age of information (AoI). In particular, differently from previous studies,
our analysis accounts for the fact that the number of terminals contending the
channel may vary over time, as a function of the duration of the previous
contention period. The stability of the protocol is analyzed via a drift
analysis, which allows us to determine the presence of stable and unstable
equilibrium points. We also provide an exact
characterization of the AoI performance, through which we determine the impact of some key
protocol parameters, such as the maximum length of the contention period, on the
average AoI. Specifically, we show that configurations of parameters that
maximize the throughput may result in a degradation of the AoI performance.
\end{abstract}

\section{Introduction} \label{se:introduction}


\bigskip

\IEEEPARstart{W}{ireless} sensor networks (WSNs) and \Ac{IoT} systems often involve a large number of terminals that sense a physical process and report time-stamped status updates to a common receiver. This scenario is relevant in, e.g., environmental monitoring, managing of connected vehicles, and asset tracking, where a primary objective is to maintain an up-to-date record of the status of an observed source. A number of performance metrics related to the notion of information freshness have recently been proposed to quantify the ability of a system to reach this goal~\cite{Yates21_JSAC,Uysal21:Semantic}. In this context, a prominent role is played by the \ac{AoI}~\cite{Kaul11_SECON,Kaul2012:INFOCOM}, which quantifies the amount of time elapsed since the newest update available at the receiver was generated at the source. \ac{AoI} has been shown to effectively capture fundamental system behaviors in a number of relevant scenarios \cite{Kellerer19_ACM,Sun20_TIT}.

Due to the possibly massive number of battery-powered, low-complexity devices
that generate traffic in a sporadic fashion, WSNs and \ac{IoT} systems typically
rely on random access protocols at the \ac{MAC} layer. In particular, random access
strategies based on variations of ALOHA \cite{Abramson:ALOHA,Roberts72:ALOHA}
are the de-facto choice in a number of commercial systems \cite{LoRa,SigFox}.
Preliminary insights on the information-freshness trade-offs that emerge in
random-access systems were derived in \cite{Yates17:AoI_SA,Yates20_ISIT}.
Specifically, these contributions illustrate that throughput and \ac{AoI} can be
optimized simultaneously under  ALOHA policies by properly tuning the channel
access probability. Further improvements in ALOHA-based protocols with feedback
were discussed in \cite{Uysal21_JSAC,Bidokhti22:TIT}. The demand for efficient
medium access control strategies for emerging massive WSNs and \ac{IoT} systems
has originated a revived interest in the design of powerful random access
protocols
\cite{DeGaudenzi07:CRDSA,Liva11:IRSA,Polyanskiy17:RA,Frolov2020,Narayanan2020,Guillaud2021,Fengler_IT}.
Among various modern random access approaches, advanced ALOHA-based schemes
\cite{Liva11:IRSA,Narayanan2012,Stefanovic12:Frameless,Paolini15:TIT_CSA,Sandgren2017,Clazzer2018}
gained popularity thanks to their remarkable performance that is achievable with
a limited complexity from a signal processing viewpoint. Such protocols allow
terminals to transmit multiple copies of their packets over time, and employ
successive interference cancellation at the receiver to resolve collisions. This
leads to significant throughput improvements, which make these solutions
excellent candidates for next-generation \ac{IoT} networks. Unfortunately,
little is known about the behavior of modern random access protocols in terms of
information freshness. The first results in this direction were presented in
\cite{Munari21_TCOM}, where the focus was on irregular repetition slotted ALOHA
\cite{Liva11:IRSA}. There, non-trivial trade-offs between throughput and average
\ac{AoI} were revealed.

Among advanced ALOHA-based random access protocols, frameless ALOHA
\cite{Stefanovic12:Frameless,Stefanovic13:RatelessAloha} emerges as the only
variant that is explicitly designed by taking into account the availability of a
feedback channel. 
While the protocols introduced in~\cite{Liva11:IRSA,Paolini15:TIT_CSA} can be placed in strong connection with the
theory of low-density parity-check codes \cite{Gal63}, frameless ALOHA operates
instead 
according to the same principle as rateless codes~\cite{Lub02}, and has emerged
as a particularly promising approach. Specifically, frameless ALOHA allows
terminals to transmit copies of their packets over a contention period whose
duration is dynamically tuned by the receiver based on the fraction of currently
unresolved collisions. A throughput analysis of frameless ALOHA was proposed in
\cite{Lazaro20_TCOM}, which is based on the Markovian analysis of the peeling
decoder for LT codes \cite{KLS04}. Both the original analysis of frameless ALOHA
\cite{Stefanovic12:Frameless,Stefanovic13:RatelessAloha} and its exact
finite-length characterization \cite{Lazaro20_TCOM} focus on the \emph{static}
behavior of the protocol, i.e., they condition the analysis on the number of
terminals becoming active in the current contention period. It follows that the
analyses of
\cite{Stefanovic12:Frameless,Stefanovic13:RatelessAloha,Lazaro20_TCOM} do not
characterize the \emph{dynamic} behavior of the protocol. It is known that
ALOHA-like protocols with feedback possess a rich dynamic behavior, which
(depending on the load and on the retransmission policy) may result in systems operating in undesirable throughput / delay regimes
\cite{kleinrock75:ALOHAPERF,kleinrock75:ALOHADYN}. 
\paragraph*{Contributions}
In this paper, we analyze the
dynamic behavior of frameless ALOHA. In contrast to previous works, which assume
a fixed number of contending users, we focus on a more general and realistic
setup in which the number of users accessing the channel may vary over time,
driven by the duration of previous contention periods. We track the dynamic
evolution of the system by means of a Markovian analysis, deriving its
stationary throughput as well as identifying the stable and  unstable  operating points of the protocol via a drift study. Moreover, we provide an exact  characterization of the average
\ac{AoI} performance of frameless ALOHA as a function of the system parameters.
\rev{From a technical perspective, the novelty of our contribution lies in the identification of a natural way to model the evolution of the system under analysis and of a convenient parametrization for the corresponding finite-state machine.}

The analysis reveals a fundamental trade-off between the \ac{AoI} and the
throughput performance of frameless ALOHA, highlighting the critical role played
by some key protocol parameters, such as the maximum length of the contention
period. 
It also shows that operating the system at maximum throughput comes at the
expense of an \ac{AoI} degradation---a trade-off that is fundamentally different
from what previously noted for traditional ALOHA strategies. We complement the analysis 
by introducing simple modifications to the frameless ALOHA
protocol, which improve the \ac{AoI}/throughput trade-off.
\paragraph*{Paper Outline}
The paper is organized as follows. In Section \ref{sec:sysModel}, we introduce the
system model, and provide basic definitions. The finite-length analysis of
the \ac{SIC} process for frameless ALOHA is outlined in Section
\ref{sec:framelessAnalysis}. The analysis of the throughput of frameless ALOHA,
accounting for the system dynamic behavior, is derived in Section
\ref{sec:throughput}. In Section \ref{sec:ageAnaylsis}, we characterize the \ac{AoI}
of frameless ALOHA and illustrate the throughput vs. \ac{AoI} trade-off via
some numerical examples. Conclusions follow in Section \ref{sec:conclusions}.

\section{System Model and Preliminaries}
\label{sec:sysModel}

We focus on a system in which \nodes\ users share a wireless channel to communicate with a common receiver (sink). Time is divided in slots of fixed duration, equal to the length of a packet, and all terminals are  slot-synchronous. The medium is shared among all users following a grant-free approach, and a collision channel model is assumed.
Specifically, the transmission of two or more packets over a slot leads to a destructive collision, which prevents immediate retrieval of all colliding packets at the sink.
On the contrary, packets sent over \emph{singleton slots} are always decoded correctly.

Channel access is regulated by the frameless ALOHA protocol~\cite{Stefanovic12:Frameless}, which operates in successive \acp{CP} of not necessarily equal length.
The receiver initiates a new \ac{CP} by broadcasting a beacon, whose duration is
considered negligible throughout our analysis. After this, every user with data
to send, attempts transmission of its packet over each subsequent slot with probability~\pTx, potentially sending multiple copies of the same packet over the \ac{CP}.
Conversely, users that do not have a packet to send at the time of beacon
reception, refrain from accessing the channel for the whole duration of the \ac{CP}.
The procedure continues until a new beacon sent by the sink notifies the end of the current \ac{CP} and the start of the next one.

At the receiver side, the decoding of a packet over a singleton slot triggers \ac{SIC}.
Specifically, the interference contribution of all the copies of the retrieved packet is removed, possibly leading to new singleton slots and thus to the decoding of previously collided packets. Note that, in order to implement this procedure, the sink needs to know  the position of all the replicas of a packet. This can be achieved, for instance, by using a hash function of the payload as seed for a pseudo-random generator, used by the transmitter to determine the slots of the \ac{CP} over which to transmit. Upon decoding the payload, the sink becomes thus aware of all the slots occupied by the user, effectively allowing the removal of the interference of that user throughout the \ac{CP}.

\begin{figure*}
  \centering
  \includegraphics[width=.72\textwidth]{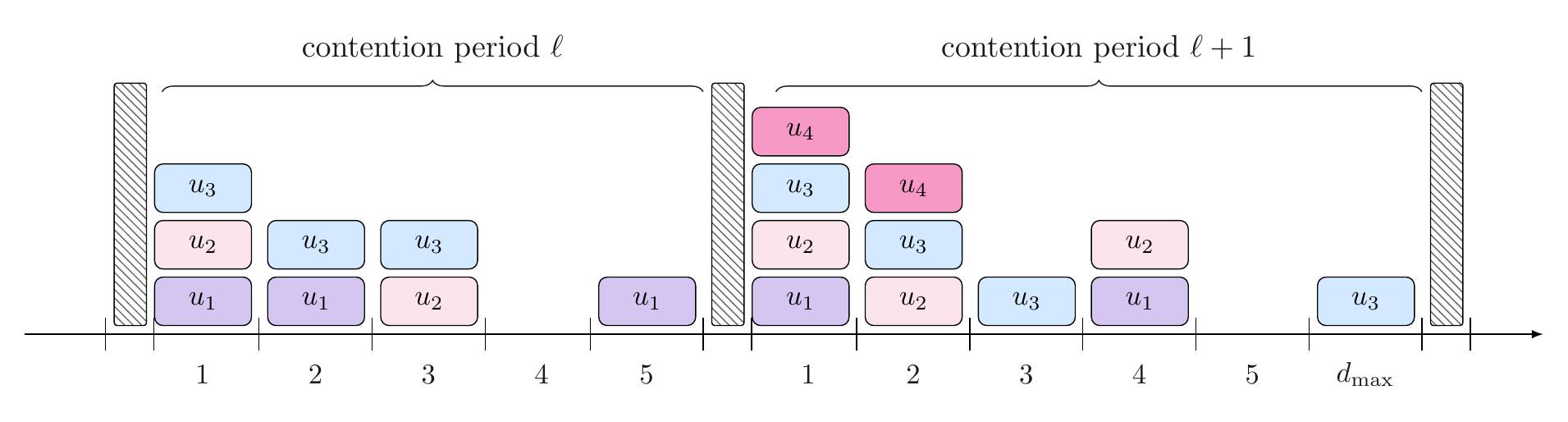}
  \caption{Example of operations for frameless ALOHA over two successive \acp{CP}.
  We assume $\nodes=4$ users in the system and a maximum contention duration of $\maxFrame=6$ slots.
Within the $\ell$-th \ac{CP}, only three users are active.
The receiver decodes the first packet in slot $5$, retrieving the status update of user $u_1$.
By removing its interference contribution from slot $2$, the sink can then decode the packet of user $u_3$. Finally, after removing the interference caused by user $3$, the sink can also obtain the packet of user $2$. Having decoded all users, the sink sends a new beacon at the end of slot $5$, initiating the next \ac{CP}. All four users attempt transmission. The first decoding occurs at slot $3$, leading to the retrieval of $u_3$. The removal of such packet from slot $2$, however, does not resolve completely the existing collision, and \ac{SIC} stops. The situation does not change after slot $4$ (collision not involving $u_3$), slot $5$ (idle), or slot $6$, which contains the transmission of a resolved user, and the receiver terminates the \ac{CP} as the maximum number of slots has been reached, even if some users (i.e., $u_1$, $u_2$ and $u_4$) have not been decoded.
Note that the first slot of each \ac{CP} is used by all active users to send a packet,
allowing the sink to infer when complete decoding has occurred (see Section~\ref{sec:operationalDetails}).}
  \label{fig:framelessTimeline}
\end{figure*}

The receiver proceeds with this operation mode on a slot-by-slot basis, and
terminates the \ac{CP} when either all transmitting users have been decoded or a
maximum number \maxFrame~of slots has been reached. Details on how the sink can
determine whether all users have been decoded will be presented in Section~\ref{sec:operationalDetails}. An example of the frameless-ALOHA operations is discussed in Fig.~\ref{fig:framelessTimeline}.

As to traffic profile, we assume every user to independently generate a new packet over each slot with probability~\pAct. This packet is stored in a one-packet-sized buffer for later delivery.
A pre-emption policy with replacement in waiting is implemented, so that, at any given time instant,
a user either has one packet to send (the last generated one) or has an empty buffer.
Accordingly, a user will attempt transmission over a \ac{CP} only if it has generated at least one packet over the previous \ac{CP}. Assuming this lasted for $\cprv$ slots, an arbitrary user has then a packet to transmit with probability
\begin{equation}
    \pActGivenL := 1-(1-\pAct)^{\cprv}.
    \label{eq:pActGivenL}
\end{equation}
All copies of the packet sent by each user during a \ac{CP} are marked with a common time stamp, set to the start time of the \ac{CP}.
Finally, no retransmissions are considered: if a packet is not decoded during the \ac{CP} it is sent over, it is simply discarded. Recalling that all users are assumed to generate traffic independently, the number $\nodesRV$ of users that become active at the end of
a \ac{CP} of \cprv\ slots is thus a binomial random variable with parameters $(\nodes,\pActGivenL)$.
In the remainder of the paper, we shall denote its \ac{PMF} as
\begin{align}
  P_{\nodesRV|\cpRV}(\nodesrv|\cprv) := \binom{\nodes}{\nodesrv} \pActGivenL^{\nodesrv} \, (1-\pActGivenL)^{\nodes-\nodesrv}.
  \label{eq:pUGivenD}
\end{align}

\begin{figure}
    \centering
    \includegraphics[width=.5\columnwidth]{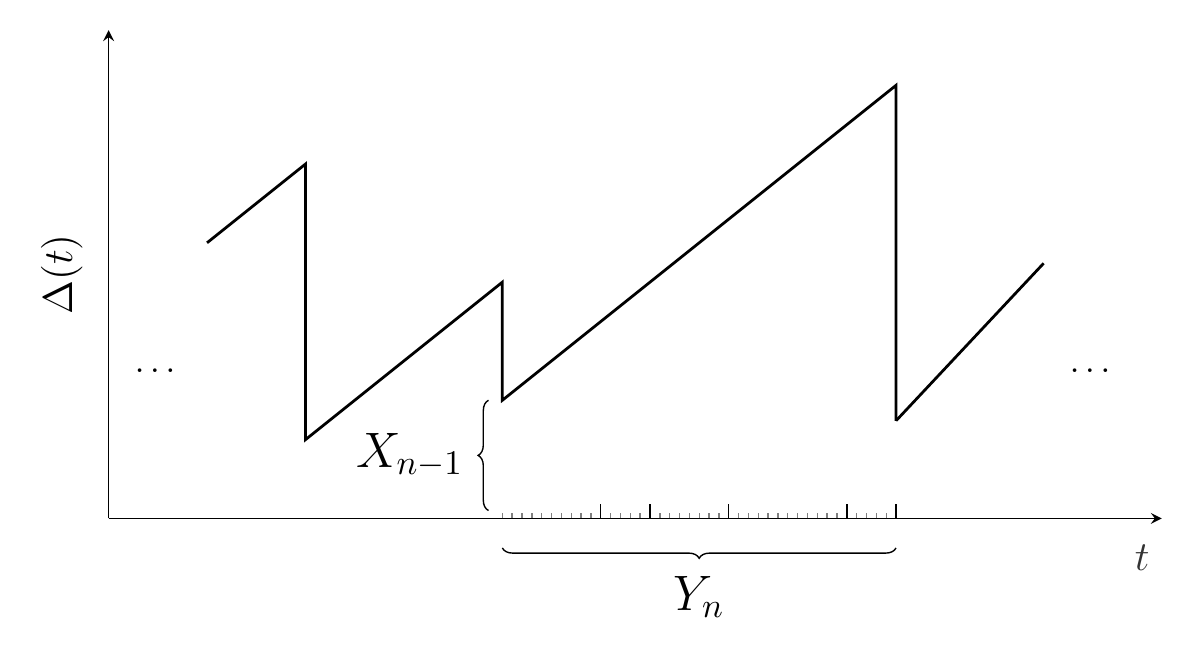}
    \caption{Evolution over time of the \ac{AoI} $\AoI(t)$ for a generic user.
    In the plot, $\interUpdate_n$ denotes the duration (in slots) of the $n$-th inter-update period, possibly composed of multiple \acp{CP}. The value at which the \ac{AoI} is reset upon reception of the node's update that starts the inter-update period is denoted by $X_{n-1}$, which in our case corresponds to the duration of the \ac{CP} in which the previous update was received.  An in-depth discussion of these quantities will be presented in Section~\ref{sec:ageAnaylsis}.
}
    \label{fig:aoiTimeline}
\end{figure}

In this paper, we are interested in evaluating the ability of the system to
maintain an up-to-date record of the state of each user at the sink. To this
aim, we consider the \ac{AoI} $\AoI(t)$ of a generic user,
\begin{equation}
\AoI(t) := t - \sigma(t)
\end{equation}
where $\sigma(t)$ is the time stamp of the last update received by the sink from the user of interest as of time $t$.  The metric grows linearly over time, and drops each time the receiver successfully decodes a packet from the user under observation.
For simplicity, we will assume that these refreshes take place at the end of the \ac{CP} over which the status update was received, i.e., we do not track the exact slot in which the corresponding packet was decoded.\footnote{As will be clarified in
Section~\ref{sec:ageAnaylsis}, this assumption does not change the fundamental trade-offs of interest, and the analysis can be easily adapted to account for this additional factor.}
This yields the saw-tooth profile exemplified in Fig.~\ref{fig:aoiTimeline}. In the remainder, we will focus on the \emph{average \acf{AoI}} \avgAoI\ \cite{Yates19_TIT}, defined as
\begin{align}
  \avgAoI := \limsup_{t\rightarrow \infty} \frac{1}{t} \int_{0}^t \AoI(\tau) d\tau.
\end{align}

\rev{To complement our study, we also evaluate the performance of the protocol in terms of \emph{expected throughput}, defined as the average number of decoded packets per slot \cite{Lazaro20_TCOM}, and denoted by \tru. Furthermore, to explore the dynamic behavior of the contention over time, we resort to a drift analysis, akin to the one often employed in the study of ALOHA systems~\cite{Gallager:DataNetworks}. Specifically, denote by \nodesRVl\ the number of users contending over the $\ell$-th CP. We shall characterize the drift for this quantity, defined as the average difference between the number of users contending over the next \ac{CP} and the number of users contending over the current \ac{CP}, given that \nodesrv\ users contended over the current \ac{CP}. 
  Let $\drift(u)$ denote the drift; we have
\begin{align}
  \drift(u) := \expOp\left[ \, \nodesRVll - \nodesRVl \,|\, \nodesRVl = \nodesrv
  \, \right].
  \label{eq:drift_definition}
\end{align}} 

\subsection{Operational Details}
\label{sec:operationalDetails}

We next describe some operational details of the protocol that will be relevant for the subsequent analyses.
At the end of each slot, the sink attempts to decode as many users as possible, canceling also their interference.
When no more users can be decoded, i.e., when the contention contains no more singleton slots, the receiver decides
whether to terminate the \ac{CP} or not.
Specifically, the \ac{CP} is concluded only if all active users have been
decoded, or, alternatively,
if a maximum number of slots has elapsed since the beginning of the contention.
Note that, without further assumptions, it is in general not possible for the sink to determine whether all active users have been decoded,
since the sink cannot discriminate between inactive users, who do not have a packet to transmit,
and active users, who do have a packet to transmit, but have not transmitted their packet (yet) since the beginning of the \ac{CP}.

To allow the sink to determine whether all active users have been decoded, we set the slot access probability to $1$ in the first slot of every contention period.
This implies that all active users will transmit their packet in the first slot.
Furthermore, we make the reasonable assumption that the receiver can distinguish among {empty slots}, {singleton slots} containing exactly one packet, and \emph{collided slots} containing two or more packets.
Under this assumption, the sink can use the first slot of every \ac{CP} to determine whether all active users have been decoded or not.
In particular, after canceling the interference from a decoded user, the sink
can check whether the first slot becomes
empty to infer whether there are no more undecoded active users and the \ac{CP} can be terminated.
This strategy allows the receiver also to detect empty \acp{CP}.
Indeed, these \acp{CP} are characterized by an empty initial slot.
Note that the minimum \ac{CP} duration in our setting is one slot, reached when either no users or a single user have data to transmit.

We emphasize that more realistic and sophisticated methods may be devised to estimate the number of active users in the \ac{CP}, as discussed for instance in~\cite{Stefanovic13:RatelessAloha}. For the purpose of the analysis provided in this paper, the proposed technique suffices, in the sense that it provides a simple model for the cost (i.e., overhead) required for the estimation of the number of active users.

\subsection{Notation}
In the remainder of the paper, we denote a discrete r.v. and its realization
using upper-case and lower-case letters such as $X$ and $x$, respectively,
whereas the \ac{PMF} of a random variable $X$ is indicated by $P_X(x)$. The
conditional \ac{PMF} of $X$ given $Y$ is denoted as $P_{X|Y}(x|y)$. We further
write the state of a homogeneous, discrete-time Markov chain at time $\ell$ as $X^{(\ell)}$,
and denote its one-step transition probability from state $i$ to state $j$ as
\begin{equation}
    \pmc{X}{i}{j} := \mathbb P \{X^{(\ell+1)} = j \, | \, X^{(\ell)} = i\}\,.
\end{equation}
In the case of bi-dimensional Markov chains, we maintain the same notation, but denote the state by means of a
two-element vector, e.g., $j = (j_1,j_2)$.

\section{Frameless ALOHA Analysis}
\label{sec:framelessAnalysis}

Following \cite{Lazaro20_TCOM}, we model the iterative \ac{SIC} process at the sink using a finite-state machine.
A state is identified by the triplet $( \unres, \collisions, \singletons)$, where $\unres$ denotes the number of unresolved users, 
$\collisions$ denotes the number of collided slots (ignoring the initial slot),  and $\singletons$ is the number of singleton slots.
We denote by $\staten{\cprv}$ the pre-decoding state, i.e., the state right after the sink observes the $\cprv$th slot within a \ac{CP} and before it tries to decode any new packets, whereas $\stateaftern{\cprv}$ denotes the post-decoding state, i.e., the state after \ac{SIC} decoding.
Note that, in the pre-decoding state, we always have $r\in\{0,1\}$, since the reception of a new slot yields at most one new singleton slot. 
Furthermore, the post-decoding state must have $\singletons=0$, since all
singleton slots result in a successful decoding operation, and the corresponding
packet as well as its replicas are removed after \ac{SIC}.
\rev{For clarity, in Example~\ref{example_pre_post} we illustrate how the pre- and post-decoding state evolve during a contention.}

\begin{example}[Pre- and post-decoding state] \label{example_pre_post}
\rev{Consider the first (leftmost) contention period shown in Fig.~\ref{fig:framelessTimeline}. 
We have that $3$ users are contending. Thus after the initial slot, we have $\staten{1}=(\unres=3,\collisions=0,\singletons= 0)$, since we have 3 undecoded users, no collided slots (according to our definition, $\collisions$ denotes the number of collided slots \emph{ignoring} the initial one), and no singleton slots. 
The post decoding state after the first slot is $\stateaftern{1}= \staten{1} = (3, 0, 0)$ since no users can be resolved.
After receiving the second slot, we have $\staten{2} = \stateaftern{2}$ = (3,1,0), since this slot is a collision.
The third slot is also a collision, thus we have $\staten{3} = \stateaftern{3}= (3,2,0)$.
Upon reception of the fourth slot, we have $\staten{4} = \stateaftern{4}= (3,2,0)$, since this slot is empty.
The fifth slot is a singleton, thus we have $\staten{5} = (3,2,1)$. In this case, this singleton slot allows the receiver to resolve all $3$ users relying on \ac{SIC}. Thus, we have  $\stateaftern{5}= (0,0,0)$.
}
\end{example}


\rev{In order to describe the decoding process}, we next provide a characterization of the conditional probability of $\stateaftern{\cprv}$ given $\staten{\cprv}$ and
of the conditional probability of $\staten{\cprv}$  given  $\stateaftern{\cprv-1}$. 

\subsection{State Initialization}
Assume that $u$ users are active. 
The state is initialized as 
\begin{equation}
\staten{1}= \begin{cases}
	( 0,  0, 0)  & \text{if } \nodesrv=0	\\
	( 1,  0, 1)  & \text{if } \nodesrv=1   \\
	( \nodesrv,  0, 0)  & \text{if } \nodesrv> 1.
\end{cases}
\label{eq:stateInit}
\end{equation}
\rev{Note that, when $u>1$ users are active, we have $c=0$ although the initial slot is collided. The reason is that $\collisions$, according to our definition, denotes the number of collided slots  ignoring the initial slot.}
\subsection{Conditional Probability of $\stateaftern{\cprv}$ Given $\staten{\cprv}$}
We next derive the conditional probability of the post-decoding state $\stateaftern{\cprv}$ given the pre-decoding state 
$\staten{\cprv} = ( \unres,  \collisions, \singletons)$.
Two cases need to be distinguished: $\singletons=0$ and $\singletons=1$.
If $\singletons=0$, the state remains unchanged since no users can be resolved. 
Hence, we have 
\begin{equation}
	\prob\{ \stateaftern{\cprv} = ( \unres',  \collisions', \singletons')|
    \staten{\cprv} = ( \unres, \collisions, 0)\}  
	= \begin{cases}
		1 & \text{if } \unres'=\unres, \,   \collisions' = \collisions, \, \singletons' = 0 \\
		0 & \text{otherwise}.
	\end{cases}
 \label{eq:condProb_triplet}
\end{equation}

Let us now focus on the case $\singletons=1$.
It is convenient to describe \ac{SIC} decoding as an iterative process in which
one user is resolved at each iteration, potentially resulting in new singleton
slots.
As a consequence, $\singletons$ may be larger than $1$ during the iterative process. 
This process is terminated when no singleton slots are available, i.e., $\singletons=0$.
To characterize the state evolution at each \ac{SIC} iteration, we use~\cite[Theorem 1]{Lazaro20_TCOM}.
This theorem, when specialized to the scenario considered here, implies that, if
the state is $(\unres,  \collisions, \singletons)$ with $\unres\geq1$ and $\singletons\geq1$, after resolving exactly one user, 
the state becomes $(\unres-1,  \collisions-\bu, \singletons-\au+\bu+\buall)$ with probability
\begin{equation}
I_\unres(\buall) \binom{\collisions}{\bu} \qu^\bu (1-\qu)^{\collisions-\bu} \binom{\singletons-1}{\au-1} \left( \frac{1}{\unres}\right)^{\au-1} \left(1- \frac{1}{\unres}\right)^{\singletons-\au}
\label{eq:recursion_2}
\end{equation}
for $\buall \in \{0,1\}$, $\bu \in \{0,\dots,\collisions\}$ , $ \au \in \{1,
\dots, \singletons-1\}$ and $\au -\bu -\buall \leq \singletons$, 
where 
\begin{equation}
I_\unres(\buall)=\begin{cases}
	1, & \text{if } \unres\neq2, \, \buall=0 \\
	1, & \text{if } \unres=2, \, \buall=1 \\
	0, & \text{otherwise}
\end{cases}
\end{equation}
and with 
\begin{equation}
\displaystyle \qu =\frac{\sum\limits_{k=2}^{\nodesrv-\unres+2} \Lambda_k k (k-1) \frac{1}{\nodesrv} \frac{\unres-1}{\nodesrv-1} \frac{\binom{\nodesrv-\unres}{k-2}}{\binom{\nodesrv-2}{k-2}}} {1 - \sum\limits_{k=1}^{\nodesrv-\unres+1} \Lambda_k \unres \frac{\binom{\nodesrv-\unres}{k-1}}{\binom{\nodesrv}{k}} - \sum\limits_{k=0}^{\nodesrv-\unres} \Lambda_k  \frac{\binom{\nodesrv-\unres}{k}}{\binom{\nodesrv}{k}} }
\end{equation} 
where $\Lambda_k=\binom{\nodesrv}{k} \pTx^k (1-\pTx)^{\nodesrv-k}$.
Here, $i$ accounts for the number of singleton slots that become empty, $j$ accounts for the number of collided slots (ignoring the initial slot) that become singletons, and $a$ takes value $1$ when the initial slot becomes a singleton and $0$ otherwise.

To derive the desired conditional probability $\prob\{ \stateaftern{\cprv} = ( \unres', \collisions', 0)| \staten{\cprv} = ( \unres,  \collisions, 1)\}$ 
for all values of $\unres'$, and $\collisions'$, we apply the result just stated iteratively, stopping when we reach a state with no singleton slots.

\subsection{Contention Period Termination}
The \ac{CP} is terminated after $\cprv<\maxFrame$ slots only if all $\nodesrv$
active users are resolved, i.e., only if the post-decoding state is
$\stateaftern{\cprv} =  (0, 0, 0)$.\footnote{Note that, although the decoder
    cannot track $\unres$, it can use the initial slot to verify whether 
$\unres=0$.}
However, when $\cprv=\maxFrame$, the \ac{CP} is terminated, no matter what the value of $\stateaftern{\maxFrame}$ is.

\subsection{Conditional Probability of $\staten{\cprv}$  Given  $\stateaftern{\cprv-1}$}
We now analyze how the state changes when one slot is added to the \ac{CP}.
To do so, we derive the conditional probability of the pre-decoding state $\staten{\cprv}$ given the post-decoding state
$\stateaftern{\cprv-1}=( \unres, \collisions, 0)$, for $\cprv \geq 2$. 
Three different cases must be considered, which result in $\staten{\cprv} = (
\unres, \collisions, 0)$, $\staten{\cprv} = ( \unres,  \collisions, 1)$, and
$\staten{\cprv} = ( \unres,  \collisions+1, 0)$, respectively. 
In the first case, the $\cprv$th slot contains no packet from any of the $\unres$ unresolved\footnote{We assume that  transmissions from already resolved users are canceled  immediately after the reception of the slot. 
Thus we only consider unresolved users in the pre-decoding state $\staten{\cprv}$.} users. 
This event, which occurs with probability $(1-\pTx)^{\unres}$, yields a pre-decoding state 
$\staten{\cprv} = ( \unres, \collisions, 0)$. 
Hence, we have
\begin{equation}
\prob \{ \staten{\cprv} = ( \unres, \collisions, 0)| \stateaftern{\cprv-1} = ( \unres,  \collisions, 0)\} = (1-\pTx)^{\unres}.
\label{eq:rec_a}
\end{equation}

In the second case, the $\cprv$th slot contains the packet of exactly one of the $\unres$ unresolved users.
It can then be verified that
\begin{equation}
\prob \{ \staten{\cprv} = ( \unres,  \collisions, 1)| \stateaftern{\cprv-1} = ( \unres,  \collisions, 0)\} =  \unres \pTx (1-\pTx)^{\unres-1}.
\label{eq:rec_b}
\end{equation}
Finally, in the third case, the $\cprv$th slot contains the transmission of two or more unresolved users, which yields  
\begin{equation}
\prob \{ \staten{\cprv} = ( \unres,  \collisions+1, 0)| \stateaftern{\cprv-1} = ( \unres,  \collisions, 0)\} =  
1- (1-\pTx)^{\unres} - \unres \pTx (1-\pTx)^{\unres-1}.
\label{eq:rec_c}
\end{equation}

\rev{In conclusion, we observe that, for a fixed contention period duration $d$, the state space of the dynamical system defined by the triplet $(w,c,r)$ has cardinality that is upper-bounded by $\mathsf{U}d^2$. By examination of the recursions \eqref{eq:condProb_triplet} and \eqref{eq:recursion_2}, we can see that, for fixed $d$, the evaluation of the state probabilities has a complexity that is $\mathcal{O}(\mathsf{U}^2d^4)$, i.e., the scaling is quadratic in the number of users, and polynomial in the contention period length. The recursions \eqref{eq:rec_a}, \eqref{eq:rec_b}, and \eqref{eq:rec_c}, instead, bear only a linear dependency on $d$ and on $\mathsf{U}$ (the three equation need to be evaluated for $0\leq c \leq d$ and $0\leq w \leq \mathsf{U}$). While this calculation shows that it might indeed be complex to adopt the analysis for very large user populations and large maximum contention period lengths, the polynomial complexity of the problem still allows one to study the system performance for range of parameters' values that are of interest for practical systems (few hundreds of users, with $d_\mathrm{max}$ in the order of few hundreds slots).}

\subsection{Derivation of Some Useful Quantities}
We will next use the state-transition probabilities just introduced to derive three quantities that will turn out important
for the characterization of the average \ac{AoI}.

The first quantity is the probability that the contention period is terminated after exactly $\cprv$ slots, given that
the number of active users is $\nodesrv$. We denote this quantity by $\pDGivenN( \cprv | \nodesrv)$. 
To characterize it, we need to consider three different cases.
The first one is $\cprv=1$. In this case, we have
\begin{equation}
\pDGivenN( 1 | \nodesrv) = \begin{cases}
    1 &\text{if } \nodesrv\in\{0,1\} \\
	0 & \text{otherwise}.
\end{cases}  
\end{equation}
The second case covers $\cprv\in\{2,\dots,\maxFrame-1\}$.
Recall that, in this case, the \ac{CP} is terminated only if all $\nodesrv$ active users are resolved.  
Hence, 
\begin{align}
\pDGivenN( \cprv | \nodesrv) = \prob \{ \stateaftern{\cprv}= ( 0, 0, 0)  \}.
\label{eq:pDGivenU_a}
\end{align}
The probability of the  remaining case, $\cprv=\maxFrame$, can be easily obtained as
\begin{align}
	\pDGivenN( \maxFrame | \nodesrv) 
	&= 1 - \sum_{\cprv=1 }^{\maxFrame-1} \pDGivenN( \cprv | \nodesrv).
\label{eq:pDGivenU_b}
\end{align}

The second quantity we are interested in is the conditional probability that
exactly $m$ users were decoded at the end of the \ac{CP}, given that $\nodesrv$
users were active.
We denote this quantity by $\pMGivenN(\decrv|\nodesrv)$. 
To characterize it, we must distinguish two cases: $m<\nodesrv$ and $m=\nodesrv$. 
When $m<\nodesrv$, since not all users were resolved, the \ac{CP} was terminated after $\maxFrame$ slots. 
Hence, we have
\begin{equation}
\pMGivenN(\decrv|\nodesrv)=  \sum_{\collisions=1}^{\maxFrame-1} \prob \{ \stateaftern{\maxFrame}= ( \nodesrv-m, \collisions, 0) \}.
\end{equation}
We consider now the case $m=\nodesrv$.
To obtain $\pMGivenN(\nodesrv|\nodesrv)$, we need to add the probabilities of all post-decoding states in which all users are decoded:
\begin{equation}
\pMGivenN(\nodesrv|\nodesrv) = \sum_{\cprv=1}^{\maxFrame} \prob \{ \stateaftern{\cprv}= ( 0, 0, 0) \}= 1 - \sum_{\decrv=0 }^{\nodesrv-1} \pMGivenN(\decrv|\nodesrv). 
\end{equation}

The third quantity of interest, which we denote by $\probmn{m}{\nodesrv}$, is the conditional probability that $m$ users are
resolved, given that $\nodesrv$ users accessed the \ac{CP} and that the \ac{CP} ran until its maximum duration $\maxFrame$. 
We can obtain $\probmn{m}{\nodesrv}$ by summing  the probabilities of all post-decoding states $\stateaftern{\maxFrame}$
in which exactly $\nodesrv-m$ active users are unresolved, and then normalizing by the sum of the probabilities of all states $\stateaftern{\maxFrame}$:
\begin{align}
\probmn{m}{\nodesrv} = \frac{\sum_{\collisions=1}^{\maxFrame-1} \prob \{
\stateaftern{\maxFrame}= ( \nodesrv-m,  \collisions, 0) \}
}{\sum_{\unres=2}^{\nodesrv}
\sum_{\collisions=1}^{\maxFrame=1} \prob \{ \stateaftern{\maxFrame}= ( \unres, \collisions, 0) \}}.
\label{eq:probDecAtMaxFrame}
\end{align}


\section{Throughput Performance}
\label{sec:throughput}

We provide in this section an analysis of the stationary throughput achievable
with the frameless ALOHA protocol, which will turn out useful for the
characterization of the \ac{AoI}. Previous works, e.g.,
\cite{Stefanovic12:Frameless,Lazaro20_TCOM,Stefanovic13:RatelessAloha,Stefanovic13_ICC},
have studied the protocol behavior either over a single \ac{CP}, or under the
assumption that the number of contending terminals is fixed.  For this scenario,
the number of packets that can be decoded under an optimized access probability
has been derived. The setting under consideration in this paper, however, is
characterized by a richer dynamic, since the number of users accessing the
channel, and thus the level of contention, may vary over time.  To appreciate
this aspect, observe how, for instance, a long \ac{CP} increases the probability
for more users to generate at least one packet over its duration.  This leads to
a harsher contention over the successive period, which, in turn, is likely to
last longer. Similarly, contentions resolved in few slots will instead drive the
system on average towards shorter and less loaded \acp{CP}.

To capture the impact on throughput of this non-trivial evolution, we start by
focusing on the homogeneous Markov processes \cpRVl\ and \nodesRVl, tracking the
duration of the $\ell$-th \ac{CP} and the number of users contending over it,
respectively. Let us first consider the former, which takes values in the set
$\{1,\dots,\maxFrame\}$. Recalling that the duration of the $(\ell+1)-$th \ac{CP}
is driven by the number of users contending over it, we compute the transition
probabilities for the Markov process as
\begin{align}
 p_{\cpRV}(i,j) &= \sum_{\nodesrv=0}^\nodes \pr\{\cpRVll=j \,|\, \nodesRVll = \nodesrv\}      \cdot \pr\{\nodesRVll = \nodesrv \,|\, \cpRVl = i\}\\
 &=\sum_{\nodesrv=0}^\nodes P_{\cpRV | \nodesRV} (j|\nodesrv) \, P_{\nodesRV|\cpRV}(\nodesrv|i)
 \label{eq:transMC_D}
\end{align}
where $P_{\nodesRV|\cpRV}$ is given in~\eqref{eq:pUGivenD}, and $P_{\cpRV |
\nodesRV}$ in~\eqref{eq:pDGivenU_a} and~\eqref{eq:pDGivenU_b}. 
Similarly, the transition probabilities for the Markov process \nodesRVl\ are
\begin{align}
 p_{\nodesRV}(i,j) &= \sum_{\cprv=1}^{\maxFrame} \pr\{\nodesRVll=j \,|\, \cpRVl = \cprv\} \cdot\pr\{\cpRVl = \cprv \,|\, \nodesRVl = i\}\\
 &=\sum_{\cprv=1}^{\maxFrame} P_{\nodesRV | \cpRV} (j|\cprv) \, P_{\cpRV|\nodesRV}(\cprv|i).
 \label{eq:transMC_U}
\end{align}
In both cases, it is easy to verify that these finite-state Markov chains are irreducible and aperiodic, and thus ergodic. 
In the remainder of the paper, we shall indicate their stationary distributions, derived by solving the corresponding balance equations, as $\statDistCP(\cprv)$ and $\statDistNodes(\nodesrv)$, respectively.

Let us now denote by \decRVl\ the number of successfully decoded users over the $\ell$-th \ac{CP}. \rev{Following this notation, the system throughput \tru, i.e.,  the \emph{average} number of decoded packets per slot, can be expressed as
\begin{align}
 \tru = \lim_{t\rightarrow\infty} \frac{\frac{1}{t} \sum_{\ell=1}^t \decRVl}{\frac{1}{t} \sum_{\ell=1}^t \cpRVl}.
 \label{eq:truDef}
\end{align}}
Observing that
\begin{align}
\pr\{\decRVl = \decrv\} = \sum_{\nodesrv=0}^\nodes P_{\decRV|\nodesRV}(\decrv|\nodesrv) \, \pr\{\nodesRVl=\nodesrv\}
\end{align}
we conclude that the statistics of \decRVl\ can be directly derived from that of
the number of contending users over the corresponding \ac{CP}. Hence, this
process also admits a stationary distribution. Accordingly, both the numerator and
denominator in~\eqref{eq:truDef} admit finite limits for $t\rightarrow\infty$ by
virtue of the ergodicity of the involved chains. This allows us to compute \tru\
as the ratio of the expected values of the processes in stationary conditions as
follows:
\begin{equation}
\tru = \frac{\sum_{\decrv = 0}^\nodes \sum_{\nodesrv=0}^\nodes \decrv \, P_{\decRV|\nodesRV}(\decrv|\nodesrv) \, \statDistNodes(\nodesrv)}{\sum_{\cprv=1}^{\maxFrame} \cprv \,\statDistCP(\cprv)}.
 \label{eq:tru}
\end{equation}

\begin{figure}
  \centering
  \includegraphics[width=0.65\columnwidth]{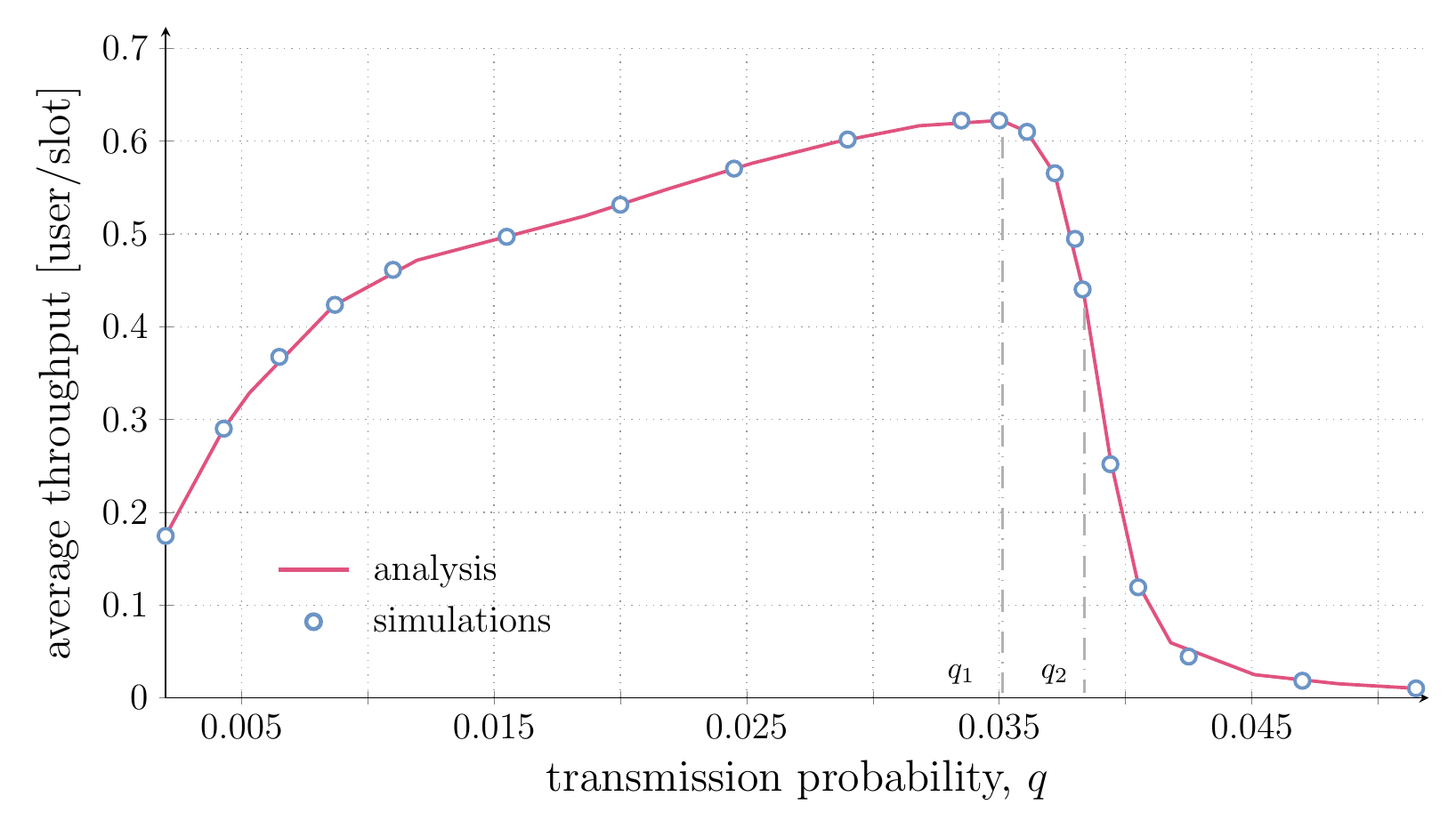}
  \caption{System throughput \tru\ vs transmission probability \pTx. A population of $\nodes=200$ users and a maximum
  \ac{CP} duration of $\maxFrame=250$ slots are considered. The packet generation probability is set so that $\pAct \nodes = 0.8$.}
  \label{fig:truVsPtx}
\end{figure}

Leaning on this result, we provide in Fig.~\ref{fig:truVsPtx} a first characterization of the behavior of the system.  
In the plot, we illustrate how the stationary throughput changes as a function of
the transmission probability \pTx, for a population of $\nodes=200$ users, and a
\ac{CP} with
maximum duration of $\maxFrame=250$ slots. The reported results
were obtained by setting the activation probability \pAct\ such that the average
number of users generating a new packet over each slot, $\pAct \nodes$, equals~$0.8$. 
The solid line shows the analytical outcomes obtained by
evaluating~\eqref{eq:tru}, whereas the markers denote results of Monte Carlo simulations. For the latter, the complete protocol operations, including traffic generation, channel access and \ac{SIC} procedures over a collision channel were implemented.

The exhibited trend confirms the existence of an optimal, throughput maximizing, medium access probability.  
Indeed, too low values of \pTx\ tend to result in successful yet unnecessarily
long \acp{CP}, where many slots may remain unused. 
Conversely, when users become
too aggressive in their transmission policies, collisions become predominant,
leading to the sharp decrease in throughput which is typically observed in
grant-free schemes that resort to \ac{SIC}~\cite{Liva11:IRSA}.

\begin{figure}
  \centering
  \includegraphics[width=0.65\columnwidth]{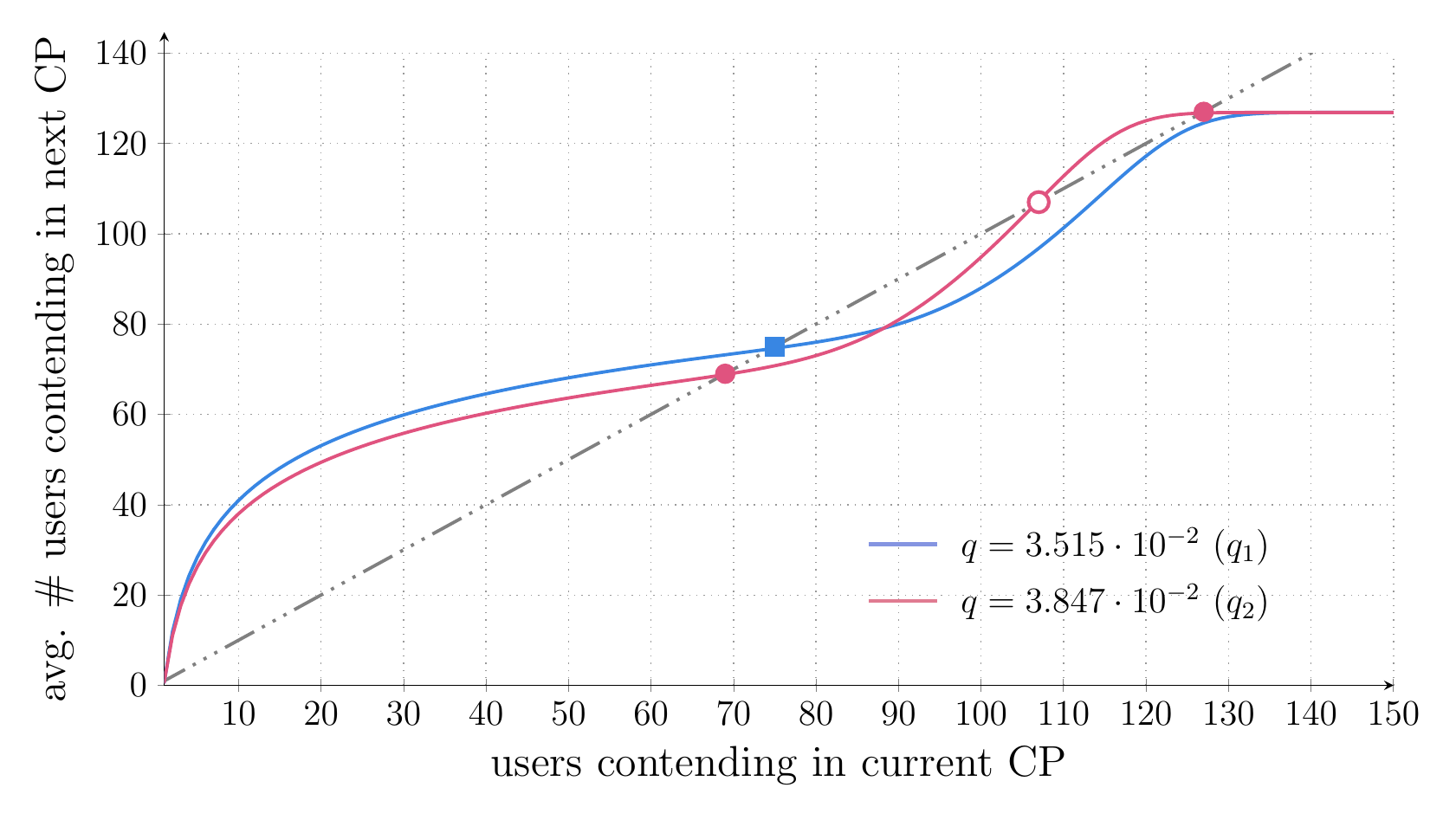}
  \caption{Graphical illustration of the drift analysis for the system. The plot reports the average number of users expected to access the channel in the next \ac{CP} as a function of the number of users contending in the current \ac{CP} (i.e., $\drift(\nodesrv)$ vs. \nodesrv). A population of $\nodes=200$ users and a maximum \ac{CP} duration of $\maxFrame=250$ slots are considered. The packet generation probability is set so that $\pAct \nodes = 0.8$. Two values of \pTx\ are studied, corresponding to the operating points highlighted in Fig.~\ref{fig:truVsPtx}. Filled markers denote stable equilibrium points, whereas the empty marker for the case $q=q_2$ indicates an unstable equilibrium. We note that, in both configurations, the curves saturate to a maximum value of $\nodes(1- (1-\pAct)^{\maxFrame}) \simeq 126$, corresponding to the number of users that become active for $\pAct\nodes = 0.8$ when the maximum \ac{CP} duration is undergone.}
  \label{fig:drift}
\end{figure}

We remark that the behavior just described is representative of the
\emph{average}
performance of frameless ALOHA, as captured by the throughput definition in~\eqref{eq:truDef}. 
To better appreciate the \emph{finer-grained dynamics} of the protocol, it is useful to resort to a \emph{drift analysis}. \rev{Recalling the definition in \eqref{eq:drift_definition}, 
we are interested in $\drift(u)$, i.e., the average difference between the number of users contending over the next \ac{CP} and the number of users contending over the current \ac{CP}, given that
\nodesrv\ users contended over the current \ac{CP}. Leaning on the transition probabilities for the Markov chain \nodesRVl, the quantity can conveniently be expressed as
\begin{align}
  \drift(u) = \sum_{i=0}^\nodes i \, p_{\nodesRV}(u,i) -\nodesrv
  \label{eq:drift_definition_v2}
\end{align}
and easily be computed for every \nodesrv\, by resorting to the
formulations derived in~\eqref{eq:transMC_U}.}
Interestingly, the drift provides an indication of how the system tends to evolve. 
Indeed,  when $\drift(u)<0$ fewer contending users are expected, whereas $\drift(\nodesrv)>0$ denotes a tendency to have more terminals attempting transmission in the upcoming \ac{CP}.  
In turn, conditions characterized by $\drift(\nodesrv)=0$ are referred to as \emph{equilibrium points}.
This behavior can be conveniently summarized using the diagram reported in Fig.~\ref{fig:drift}, representative of a system with $\nodes=200$ users and a maximum \ac{CP} duration of $250$ slots.\footnote{\rev{The analytical results for the stability study were also verified by means of dedicated numerical simulations. The outcomes, which indicate an excellent match,  are not reported, to avoid crowding the figure.}} 
The plot reports, for any value of \nodesrv, the average number of users contending over the next \ac{CP}, i.e., $\drift(\nodesrv) + \nodesrv$ (solid lines). 
A bisector of the plane (dashed line) is also shown, so that for any point on
the $\nodesrv$-axis, the drift corresponds to the difference between the solid and dashed curves.

\begin{figure*}[t]
  \begin{center}
  \includegraphics[width=.165\textwidth]{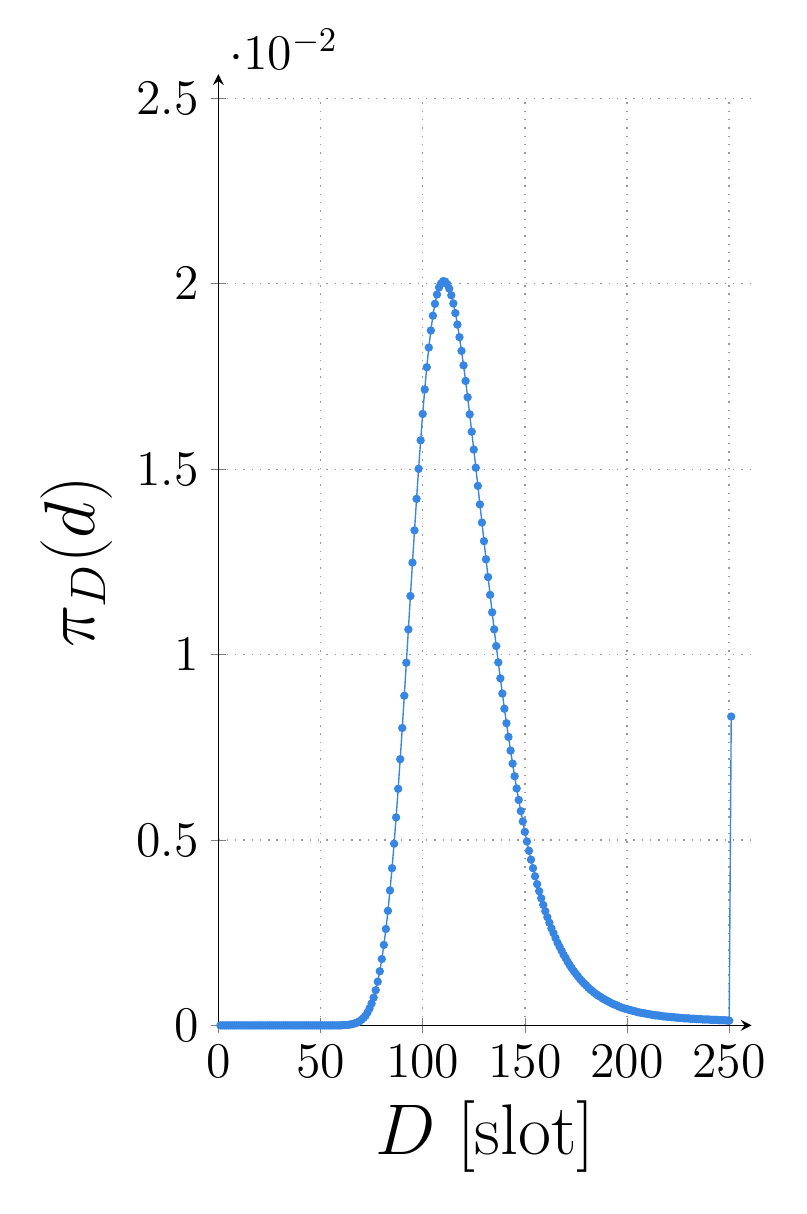}
 \hspace{-1em}
  \includegraphics[width=.165\textwidth]{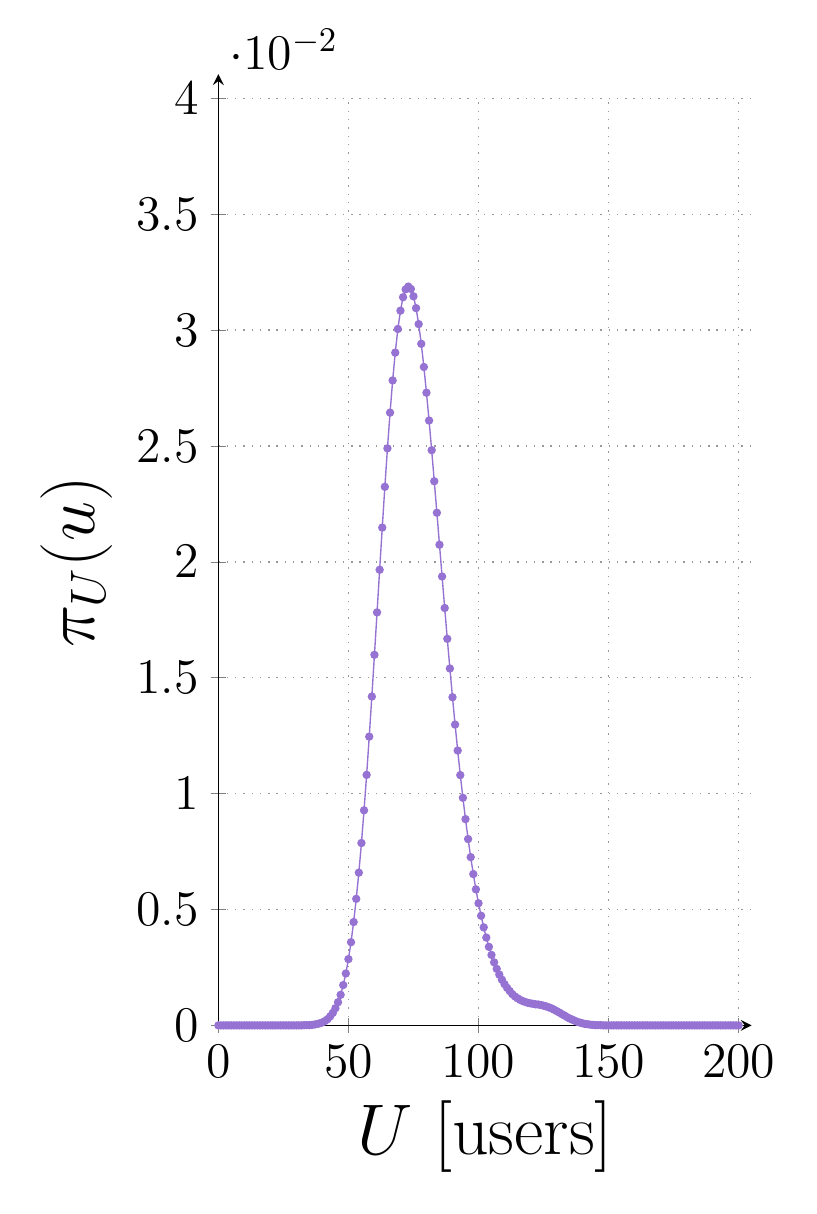}
   \hspace{-1em}
  \includegraphics[width=.165\textwidth]{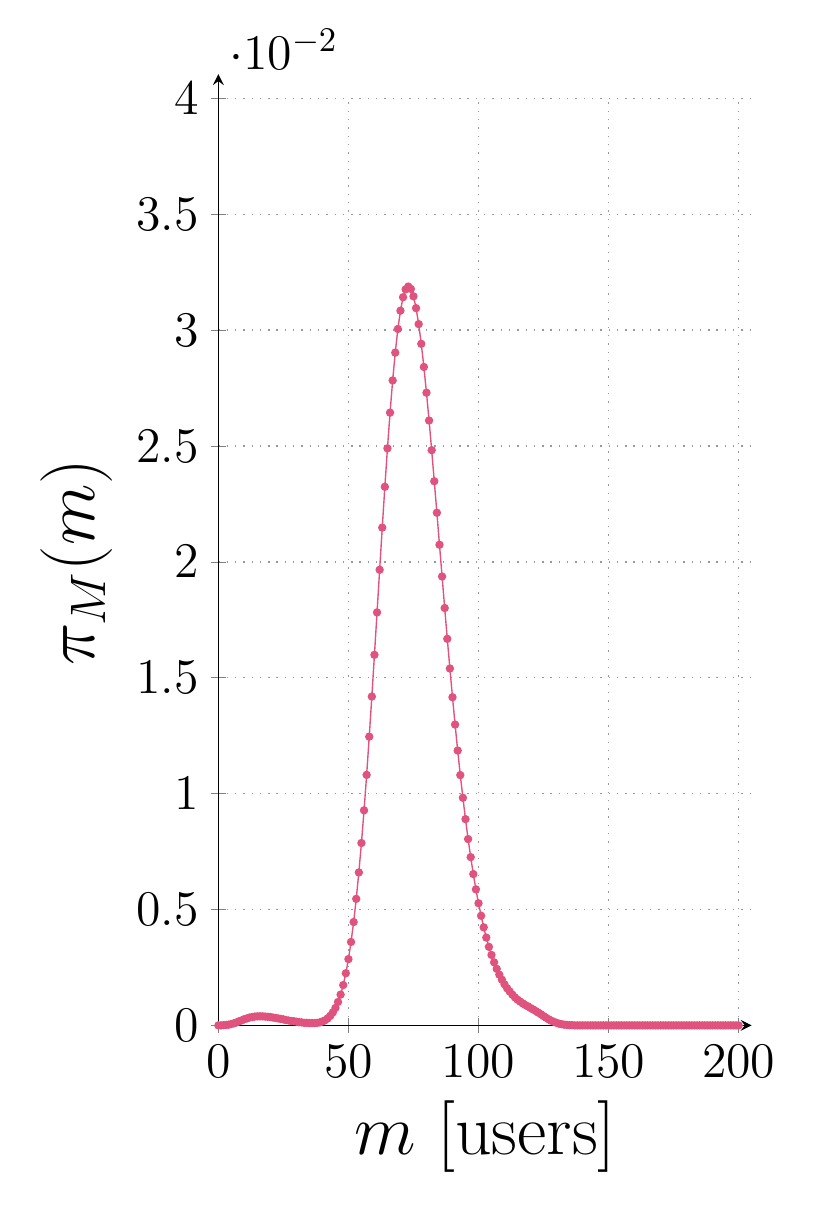}
   \includegraphics[width=.165\textwidth]{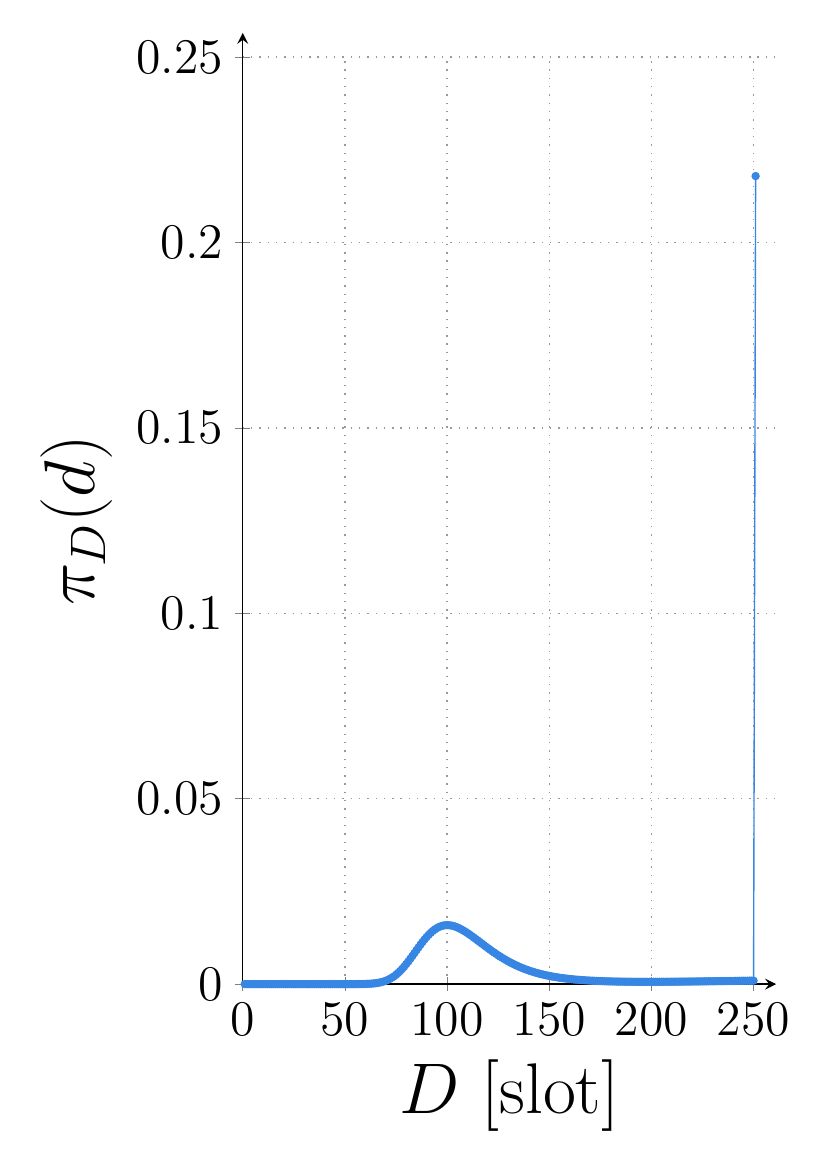}
    \hspace{-1em}
  \includegraphics[width=.165\textwidth]{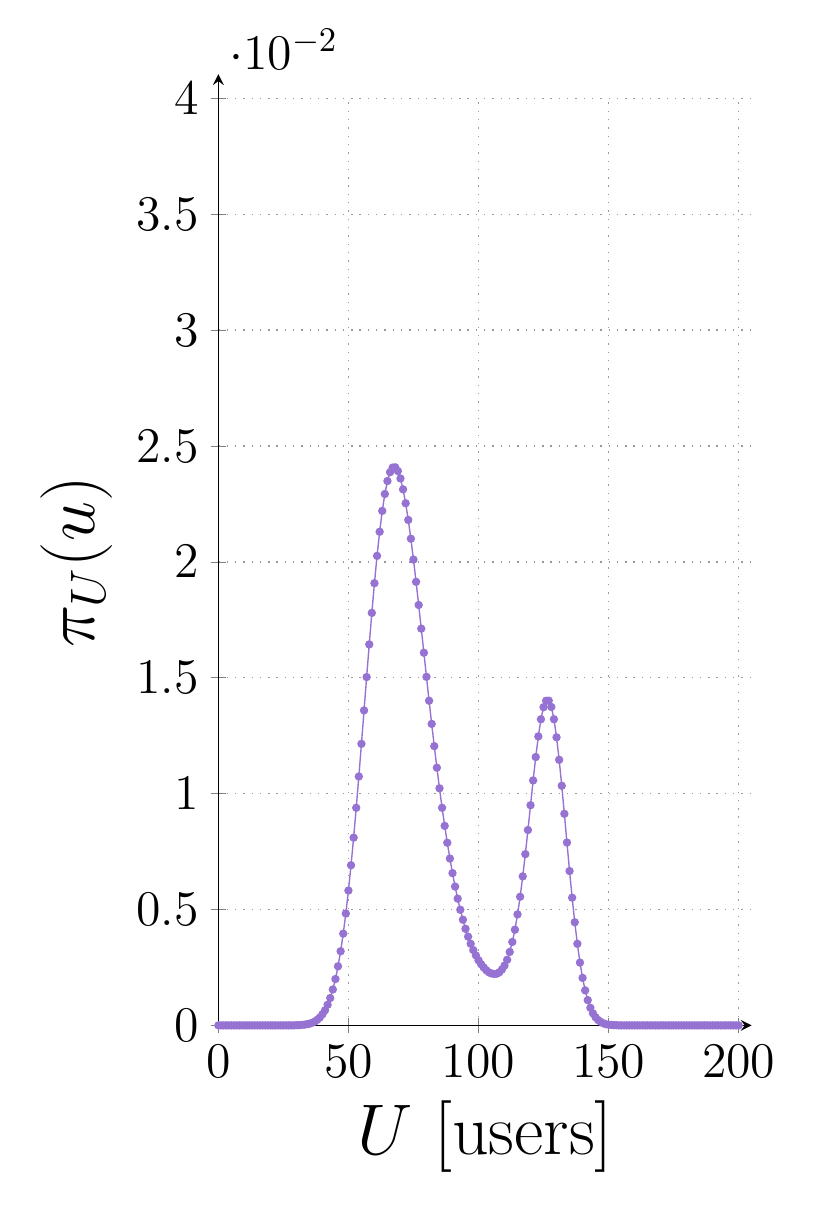}
  \hspace{-1em}
  \includegraphics[width=.165\textwidth]{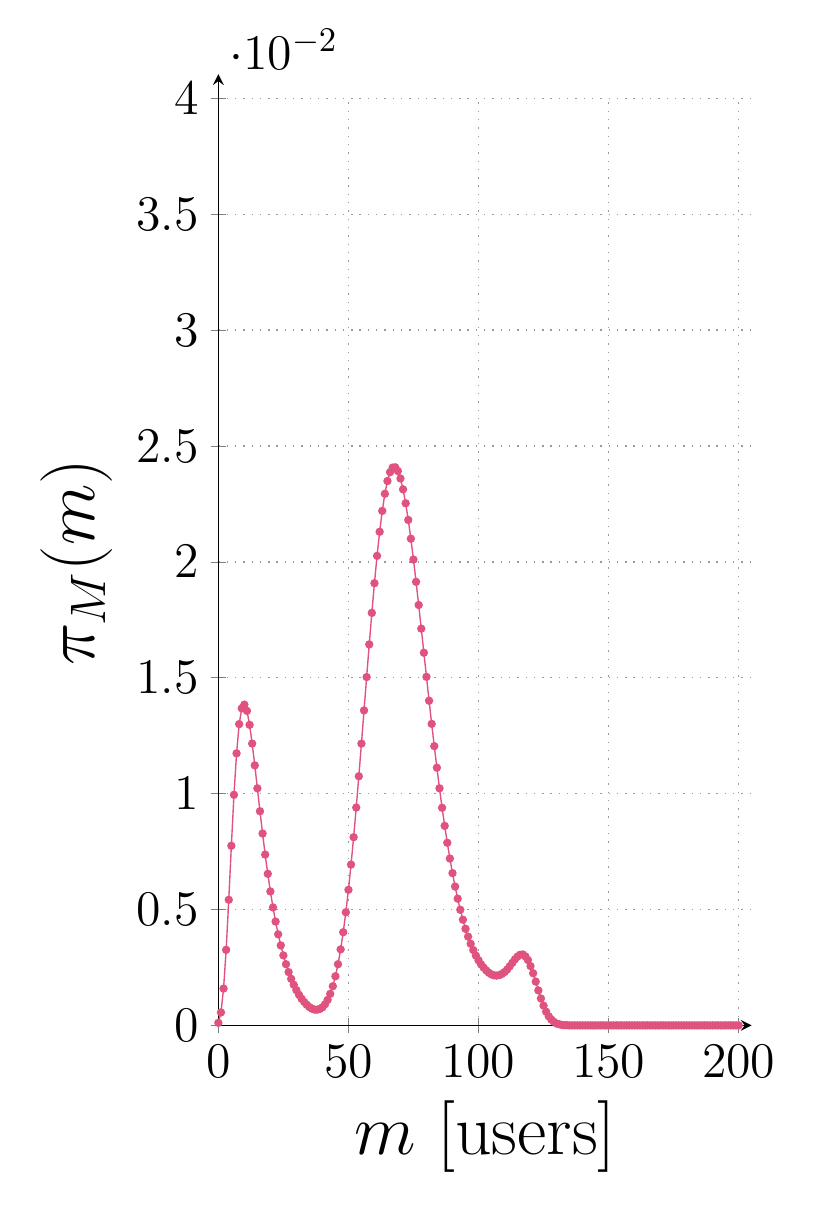}
  \end{center}
  \vspace{-.6em}
  \footnotesize
  \hspace{15em} $(\pTx=q_1)$ \hspace{23em} $(\pTx=q_2)$
  \caption{Stationary distribution of the \ac{CP} duration, $\statDistCP(\cprv)$, number of contending users per CP $\pi_{\nodesRV}(\nodesrv)$, and number of decoded users per \ac{CP}, $\pi_{\decRV}(\decrv)$, for the two values of transmission probability $\pTx_1$ and $\pTx_2$. In all cases, results were generated considering $\nodes=200$, $\maxFrame=250$, and $\pAct \nodes = 0.8$.}
  \label{fig:statDistrib}
\end{figure*}

Consider first the blue curve, obtained for a transmission probability $\pTx =
\pTx_1 = 0.03515$. Such value, also highlighted in Fig.~\ref{fig:truVsPtx},
maximizes the average throughput achieved by the protocol for the \maxFrame\
under study. In this configuration, the drift plot pinpoints the existence of a
single equilibrium point, indicated by the square marker and attained for
$\nodesrv\simeq 75$. This behavior is confirmed by the leftmost results in
Fig.~\ref{fig:statDistrib}, reporting the stationary distributions for the
\ac{CP} duration (\statDistCP), number of contending users (\statDistNodes), and
number of decoded users over a \ac{CP} (\statDistDec). When $\pTx$ is chosen so
as to provide the optimal throughput, the contention level is well concentrated around a desired value, with small fluctuations in the \ac{CP} duration and number of transmitting nodes due to the random nature of the access procedures.

The situation changes drastically when the transmission probability is slightly
increased, i.e., if one sets $\pTx = \pTx_2 = 0.03847$. In this case,
Fig.~\ref{fig:drift} (purple curve) reveals the existence of three
equilibrium points.
The middle one (empty marker) is unstable, in the sense that the
system will tend to move away from it although it expresses a null
drift.\footnote{Note indeed that, for any \nodesrv\ to the left of this
equilibrium point, we have positive drift, and the system will tend to move to
more contending nodes. Similarly, for any \nodesrv\ to the right of the point,
$\drift(\nodesrv) < 0$, once again drifting away from the equilibrium.}
Conversely, the leftmost and rightmost ones (filled markers) are stable, and hint at a more
complex behavior. This can be appreciated by looking at the stationary
distributions on the right part of Fig.~\ref{fig:statDistrib}, which exhibit a
bimodal structure. In this case, the system oscillates between a desirable
equilibrium, granting a good throughput and characterized by \ac{CP} durations
and number of contending users similar to the ones observed for $\pTx=\pTx_1$,
and a detrimental one. In the latter condition, \acp{CP} of the maximum
duration are experienced, triggering more contention (rightmost peak of
\statDistNodes) and a lower success rate (leftmost peak of \statDistDec). 

From this standpoint, two remarks are in order. First, when operating in the
detrimental
configuration, the protocol achieves poor throughput, triggering the reduction
in the average performance observable in Fig.~\ref{fig:truVsPtx}. Second, and
perhaps more relevant from a practical perspective, the system may require a
long time before returning to the throughput-efficient equilibrium point, once it has reached
the undesired equilibrium point. One way to measure this elapsed time is to
count the
number of slots elapsed on average until the system returns to a favorable configuration
after having experienced a \ac{CP} of the maximum duration. 
In the setting under
study, we declare return to a desired equilibrium as soon as a contention is
terminated after less than $150$ slots. This value was chosen as representative
as it is one of the largest \ac{CP} durations still lying close to the first peak
of the distribution \statDistCP\ (see Fig.~\ref{fig:statDistrib}). For the
considered parameters, this transition takes approximately $14200$
slots.\footnote{The reported value was obtained by means of a first step
analysis of the involved Markov processes. The analytical details, omitted here,
follow the same methodology that will be presented in depth in Section~\ref{sec:ageAnaylsis} in the context of the \ac{AoI} study.} In other
words, a period corresponding to roughly $150$ \acp{CP} of $100$ slots---a typical slot duration for the case in which the system in the desirable equilibrium point---are wasted because the system is stuck in the undesired equilibrium point. \rev{In this situation, a reset may be required to shorten the time spent in highly inefficient conditions, with an extra cost in terms of overhead.}

\rev{To summarize, our study reveals that frameless ALOHA has a complex dynamic behavior, which calls for a careful tuning of the system parameters. In this perspective, the presented analysis offers a useful tool for an initial system design.}

\section{Average Age of Information}
\label{sec:ageAnaylsis}

We now focus on the characterization of frameless ALOHA in terms of information
freshness. To this aim, we provide first some preliminary results that will facilitate the derivation of the average \ac{AoI}.

Fix a generic user for which the \ac{AoI} is tracked, and denote by $\pUpdate(\nodesrv,\cprv)$ the conditional probability that the user delivers a status update over the current \ac{CP}, given that $\nodesrv$ users contend and that the \ac{CP} has a duration of $\cprv$ slots. Consider first the case in which the \ac{CP} is terminated prior to reaching its maximum length. Recalling the protocol operation, this condition occurs when all contending users are successfully decoded. Accordingly, $\pUpdate(\nodesrv,\cprv)$ is simply given by the probability for the node of interest to have participated to the contention, given by $\nodesrv/\nodes$. Conversely, if the \ac{CP} runs for \maxFrame\ slots, the conditional probability for the user to deliver a packet given that $\decrv$ users are
successfully decoded  can be obtained as
$(\decrv/\nodes)\decAtMaxFrame(\decrv,\nodesrv)$, as a consequence of~\eqref{eq:probDecAtMaxFrame}.
Combining these two results we then have:
\begin{align}
    \pUpdate(\nodesrv,\cprv) =
    \begin{dcases}
    \,\,\frac{\nodesrv}{\nodes} & \quad \cprv < \maxFrame \\
    \,\,\sum_{\decrv=0}^\nodesrv \frac{\decrv}{\nodes} \,\, \beta(\decrv,\nodesrv) & \quad \cprv = \maxFrame.
    \end{dcases}
    \label{eq:pup}
\end{align}

Next, we introduce an ancillary Markov chain, whose state is defined as \mbox{$\ancMCl = (\cpRVl, \succRVl)$}. The
first component, which we have already discussed, characterizes the duration of the $\ell$-th \ac{CP}, whereas \succRVl\ is a binary
r.v. taking value~$1$ if an update from the user of interest has been
successfully received over the $\ell$-th \ac{CP}, and taking value $0$ otherwise. 
Consider now the probability for the chain to transition from state $(j,\succrv)$ to state
$(\cprv,1)$. By definition, this event occurs when the current \ac{CP} has duration $\cprv$ slots, and the user
delivers an update. Observing that the user's success does not depend on its outcome over the previous \ac{CP}, we can
simplify the transition probability to
\begin{align}
      p_{\ancMC}((j,\succrv),(\cprv,1)) &:=\pr\{ \ancMCll = (\cprv,1) \,|\, \ancMCl = (j,\succrv) \}\\
    &= \pr\{ \succRVll = 1, \cpRVll = \cprv \,|\, \cpRVl = j \}.
\end{align}
Conditioning now on the number of users contending over the \ac{CP}, we further have
\begin{align}
  \pr\{ &\succRVll = 1, \cpRVll = \cprv \,|\, \cpRVl = j \} \\
  &= \sum_{\nodesrv=0}^\nodes \pr\{\succRVll=1 \,|\, \nodesRVll=\nodesrv, \cpRVll=d\}
  \cdot\pr\{ \nodesRVll=\nodesrv, \cpRVll=\cprv \,|\, \cpRVl = j\}\\
  &= \sum_{\nodesrv=0}^\nodes \pr\{\succRVll=1 \,|\, \nodesRVll=\nodesrv,
  \cpRVll=d\}  \cdot \pr\{\cpRVll\!\!=\cprv \,|\, \nodesRVll\!\!=\nodesrv \}
  \notag\\
  &\quad\quad\cdot \pr\{\nodesRVll\!\!=\nodesrv \,|\,\cpRVl \!\!= j\}.
\end{align}
Finally, using~\eqref{eq:pup},~\eqref{eq:pUGivenD}, \eqref{eq:pDGivenU_a}, and~\eqref{eq:pDGivenU_b}, we can write
$p_{\ancMC}((j,\succrv),(\cprv,1))$ compactly as
\begin{equation}
     p_{\ancMC}((j,\succrv),(\cprv,1)) =\sum_{\nodesrv=0}^\nodes \pUpdate(\nodesrv,\cprv) \,\,  P_{\cpRV|\nodesRV}(\cprv|\nodesrv) \,\, P_{\nodesRV|\cpRV}(\nodesrv|j).
     \label{eq:transProbZSucc}
\end{equation}

Following similar steps, we can express the transition probabilities from a generic state $(j,\succrv)$ to a state
$(\cprv,0)$ in which the user does not deliver an update as
\begin{align}
     p_{\ancMC}((j,\succrv),(\cprv,0)) =\sum_{\nodesrv=0}^\nodes (1-\pUpdate(\nodesrv,\cprv) )\,\,  P_{\cpRV|\nodesRV}(\cprv|\nodesrv) \,\, P_{\nodesRV|\cpRV}(\nodesrv|j).
     \label{eq:transProbZFail}
\end{align}
It is immediate to verify that the finite-state Markov chain $\ancMCl$ is irreducible and aperiodic, and admits thus a
stationary distribution, which we denote as $\statDistAncMC(\cprv,\succrv)$.

\subsection{Average AoI Analysis}

Let us now compute the average AoI achieved by frameless ALOHA. The metric can
be conveniently expressed in terms of the \emph{inter-update time}
$\interUpdate_n$ and the \emph{system time} $\initAoI_n$, introduced in
Fig.~\ref{fig:aoiTimeline}. 
The former quantity captures the number of slots
elapsed between two successive update deliveries from the node of interest. 
The latter denotes the time between the generation of an update and its
delivery, which, in our case, corresponds to the duration of the \ac{CP} over
which the update was decoded. With this notation, we have via standard
geometrical arguments, see, e.g., \cite[Eq.~(3)]{Yates21_JSAC},
\begin{align}
\avgAoI = \frac{\expOp\left[\initAoI_n\interUpdate_n\right] + \expOp\left[ \interUpdate_n^2  \right]/2}{\expOp\left[\interUpdate_n\right]}
\label{eq:avgAoI_basic}
\end{align}
under the assumption that $(\interUpdate_n,\initAoI_n)$ is a stationary ergodic process. In the remainder, we drop for ease of notation the index $n$ denoting a specific update delivery, and focus on the stationary behavior of the quantities of interest.

As initial remark, we observe that the \ac{PMF} $P_{\initAoI}(\initaoi)$ of the system time can be readily computed from the stationary distribution of the Markov chain $\ancMCl$. We have
\begin{equation}
  P_{\initAoI}(\initaoi) = \frac{\statDistAncMC(\initaoi,1)}{\sum_{\delta=1}^{\maxFrame} \statDistAncMC(\delta,1)}
  \label{eq:pmfDelta0}
\end{equation}
where the numerator denotes the probability for the system to be in a \ac{CP} of duration $\initaoi$ slots in which the tracked user is decoded, and the denominator is a normalization factor, capturing that we are interested only in \acp{CP} with successful updates from that user.
Secondly, note that, in the evaluation of \eqref{eq:avgAoI_basic}, we need to account for the statistical dependence between \initAoI\ and \interUpdate. In fact, the duration of the \ac{CP} over which the last update was received does influence the number of users contending on the subsequent one, impacting both the probability for the user of interest to transmit and be decoded as well as the duration of the subsequent \acp{CP}.

It is therefore convenient to compute the statistical moments in \eqref{eq:avgAoI_basic} by conditioning on the system time. Let us start by considering the term $\expOp[\initAoI\interUpdate]$, which we expand as
\begin{align}
  \expOp[\initAoI\interUpdate] = \sum_{\initaoi=1}^{\maxFrame} \initaoi \cdot \expOp[\interUpdate \,|\, \initAoI = \initaoi] \, P_{\initAoI}(\initaoi).
  \label{eq:exp_XY}
\end{align}
Without loss of generality, let us denote by $\ell=1$ the index of the first
\ac{CP} that contributed to the inter-update time being tracked. Accordingly, we
reformulate the conditional expectation in~\eqref{eq:exp_XY} considering the value of $\ancMC^{(1)}$ as
\begin{align}
  \expOp[ \interUpdate \,|\, \initAoI = \initaoi ]  &= \sum_{z} \,\expOp[ \interUpdate \,|\, \ancMC^{(1)} = z, \initAoI=\initaoi ] \,  \pr\{ \ancMC^{(1)} = z \,|\, \initAoI=\initaoi \}\\
  &=\sum_{z} \,\expOp[ \interUpdate \,|\, \ancMC^{(1)} \!\!= z] \,  \pr\{ \ancMC^{(1)} = z \,|\, \initAoI=\initaoi \}
  \label{eq:interUpdate_givenX}
\end{align}
where the summation is taken over all the possible states \mbox{$z = (\cprv,\succrv)$}, $\cprv\in\{1,\dots,\maxFrame\}$,
$\succrv\in\{0,1\}$, and~\eqref{eq:interUpdate_givenX} follows from the Markov property of the involved processes. 
Note that the factors $\pr\{\ancMC^{(1)} \!\!= z \,|\, \initAoI=\initaoi \}$ on the right-hand side of~\eqref{eq:interUpdate_givenX}
can be computed using~\eqref{eq:transProbZSucc} and~\eqref{eq:transProbZFail}.

The conditional expectation $\expOp[\interUpdate\,|\, \ancMC^{(1)} = z]$ can be
derived by resorting to a first step analysis \cite{TaylorKarlin}. To this aim, consider first the situation in which the packet from the user of interest is decoded
already in the initial \ac{CP}.
In this case, \interUpdate\ coincides with the length of the initial \ac{CP}:
\begin{align}
\expOp[\interUpdate \,|\, \ancMC^{(1)}= (\cprv,1)] = \cprv.
\label{eq:firstStep_a}
\end{align}
When $\ancMC^{(1)} = (\cprv,0) $ \
instead, the inter-update time can be computed as the sum of the durations of all \acp{CP} until the next update decoding. This can be conveniently computed by conditioning on the outcome of the first transition. Specifically,
\begin{align}
  \expOp[\interUpdate \,|\, \ancMC^{(1)}= (\cprv,0)] = \cprv + \sum_{z} \expOp[\interUpdate \,|\, \ancMC^{(1)}= z] \cdot p_{\ancMC}((\cprv,0),z)
  \label{eq:firstStep_b}
\end{align}
where the Markov property ensures that the average duration, once the transition
to state $z$ has occurred, is equal to the one that we would have by starting from such state.
Combining \eqref{eq:firstStep_a} and \eqref{eq:firstStep_b}, we obtain a
full-rank system of $\maxFrame$ equations in the \maxFrame\ unkowns
$\expOp[\interUpdate\,|\,\ancMC^{(1)}=(\cprv,0)]$. Substituting the solutions of
this system into \eqref{eq:interUpdate_givenX}, we obtain $\expOp[\interUpdate
\,|\, \initAoI = \initaoi]$, which eventually allows us to compute
$\expOp[\initAoI\interUpdate]$ via~\eqref{eq:exp_XY}.
We also note that the availability of $\expOp[\interUpdate\,|\, \initAoI
= \initaoi]$ allows us to evaluate the denominator of~\eqref{eq:avgAoI_basic}
via
\begin{align}
  \expOp\left[ \interUpdate \right] = \sum_{\initaoi=1}^{\maxFrame}  \expOp[\interUpdate \,|\, \initAoI = \initaoi] \, P_{\initAoI}(\initaoi).
  \label{eq:exp_Y}
\end{align}

The final term required to evaluate $\bar{\Delta}$ in~\eqref{eq:avgAoI_basic}
is~$\expOp[\interUpdate^2]$.
As before, we compute first the conditional expectation of
$\interUpdate^{2}$ given $\ancMCone$ via a first-step analysis. 
This yields the following system of linear equations:
\begin{equation}
   \expOp[\interUpdate^2 \,|\, \ancMC^{(1)}= (\cprv,1)] = \cprv^2
\end{equation}
and
\begin{IEEEeqnarray}{rCl}
    \expOp[\interUpdate^2 \,|\, \ancMC^{(1)}= (\cprv,0)] &=& \cprv^2 + 2\cprv
  \sum_{\ancmc} \expOp[\interUpdate \,|\, \ancMCone = \ancmc\,] \cdot
  p_{\ancMC}((\cprv,0),\ancmc) \notag\\
                                                         &&+ \sum_{\ancmc} \expOp[\interUpdate^2 \,|\, \ancMCone=\ancmc\, ] \cdot p_{\ancMC}((\cprv,0),\ancmc)
  \label{eq:firstStep_Y2}
\end{IEEEeqnarray}
%
%
%
where the terms $\expOp[\interUpdate\,|\, \ancMCone=\ancmc]$ were derived earlier. 
Solving this full-rank system of equation, we obtain the desired conditional
second-order moments of \interUpdate, from which we compute
$\expOp[\interUpdate^2]$ via
\begin{align}
  \expOp[\interUpdate^2] = \sum_{\initaoi=1}^{\maxFrame} \expOp[\interUpdate^2 \,|\, \initAoI = \initaoi] \, P_{\initAoI}(\initaoi).
  \label{eq:exp_Y2}
\end{align}
The average AoI \avgAoI\ is then obtained by simply inserting \eqref{eq:exp_XY},
\eqref{eq:exp_Y}, and \eqref{eq:exp_Y2} into \eqref{eq:avgAoI_basic}.

\begin{figure}
  \centering
  \includegraphics[width=0.65\columnwidth]{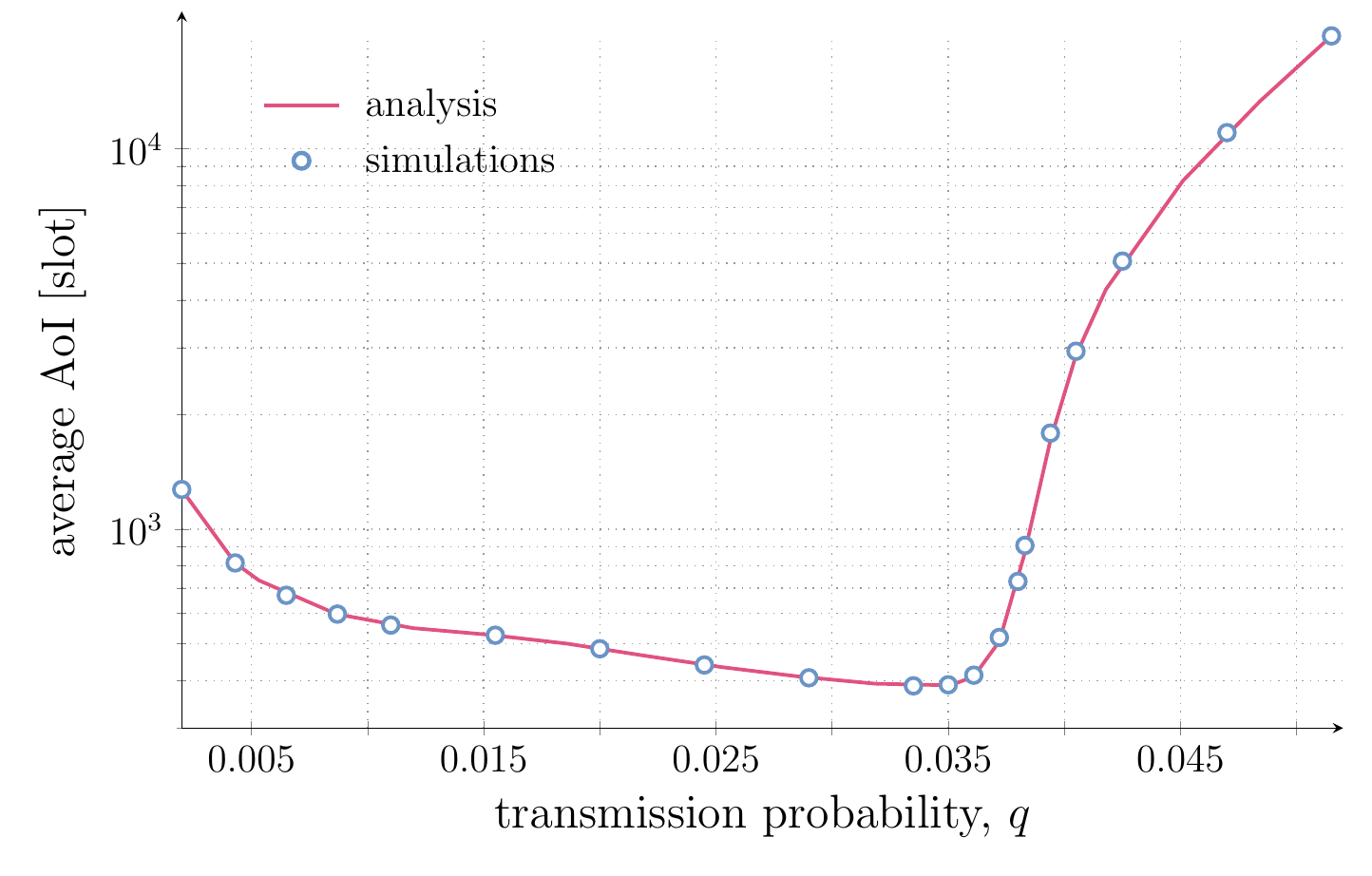}
  \caption{Average node AoI \avgAoI\ vs transmission probability \pTx. A population of $\nodes=200$ users and a maximum
  \ac{CP} duration of $\maxFrame=250$ slots are considered. The packet generation probability is set so that $\pAct \nodes = 0.8$.}
  \label{fig:aoiVsPtx}
\end{figure}

\subsection{Numerical Results}

\subsubsection{Throughput-AoI trade-off}
Leaning on the exact analysis presented so far, we illustrate the protocol behavior in terms of
information freshness in Fig.~\ref{fig:aoiVsPtx}, where we report the average
AoI as a function of the channel access probability \pTx. The same setting used
for the throughput study of Fig.~\ref{fig:truVsPtx} is considered, i.e.,
$\nodes=200$ nodes, a maximum \ac{CP} duration of $\maxFrame=250$ slots, and an
activation probability such that $\pAct \nodes = 0.8$. In the plot, the solid
lines denote analytical results, whereas markers denote the outcome of Monte Carlo simulations.
We see from the figure that too low or too high values of \pTx\ result in poor
performance in terms of \ac{AoI}. In the former case, an excessively
conservative transmission behavior is likely to result in a user missing
opportunities to deliver a status update: a user may end up not transmitting for the whole 
duration of a \ac{CP} even when a packet is available.
Conversely, when users become too aggressive, collisions dominate, hindering the
capability of the receiver to decode transmitted updates prior to reaching the
maximum contention duration.

By comparing Fig.~\ref{fig:truVsPtx} and Fig.~\ref{fig:aoiVsPtx}, we see that
the optimal operating points in terms of throughput and average \ac{AoI}
coincide. More generally, for a given traffic profile $(\nodes,\pAct)$, and a
given maximum \ac{CP} duration, there exists a value $\pTx^{*}$ of the transmission probability that jointly maximizes \tru\ and minimizes $\avgAoI$.
This outcome is common to other random access solutions under symmetric traffic
conditions, as epitomized by the inverse proportionality of \ac{AoI} and
throughput exhibited by slotted ALOHA \cite{Yates17:AoI_SA,Munari21_TCOM}.
From this standpoint, any choice of $\pTx\neq \pTx^*$ reducing the
probability to deliver a status update would also be harmful in terms of
information freshness. \rev{In the remainder, we shall refer to these optimal results as $\tru^*(\maxFrame) = \max\nolimits_{q} \tru(\pTx,\maxFrame)$ and $\AoI^*(\maxFrame) = \min\nolimits_{q} \AoI(\pTx,\maxFrame)$, explicitly pointing out that the quantities are a function of the \ac{CP} duration.}

\begin{figure}
  \centering
  \includegraphics[width=0.65\columnwidth]{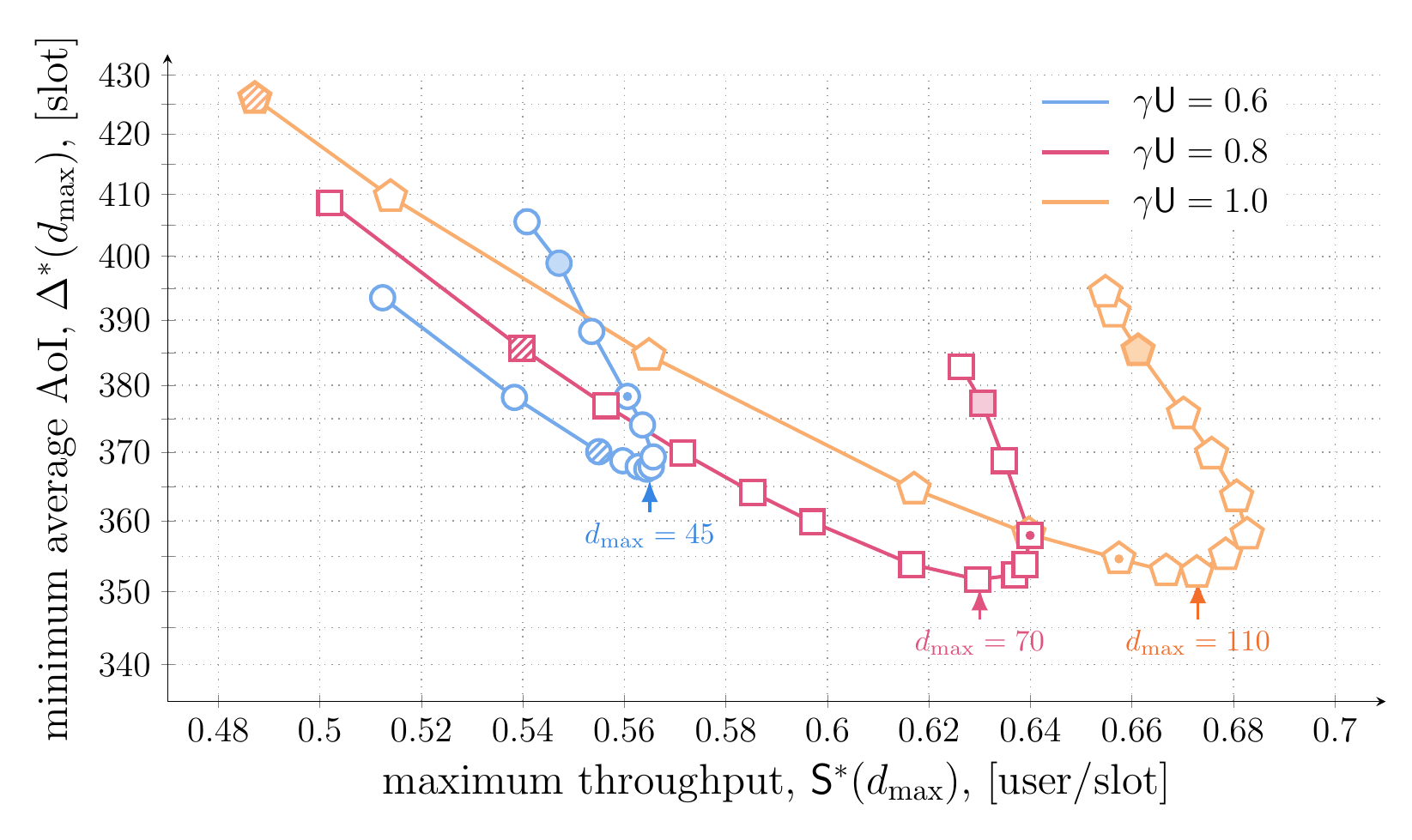}
  \caption{Maximum throughput and minimum average AoI obtained when
varying the maximum CP duration \maxFrame. Different lines correspond to results
obtained for different values of $\pAct\nodes$; in all cases $\nodes=200$. Each point is obtained considering a different value of \maxFrame, and corresponds to the maximum throughput and the minimum AoI in such conditions, i.e., $\tru^*(\maxFrame)$ and $\AoI^*(\maxFrame)$. \rev{For $\pAct\nodes=0.6$, the range $10\leq \maxFrame \leq 220$ is shown; for $\pAct\nodes = 0.8$, $20\leq \maxFrame \leq 220$; and for $\pAct\nodes=1.0$, $40\leq \maxFrame \leq 220$. The arrows point to the configuration that minimizes the AoI in each case. For reference, striped markers indicate performance for $\maxFrame=30$, markers filled with a point the case $\maxFrame=100$, and color-filled markers denote $\maxFrame=180$.}
}
  \label{fig:contour}
\end{figure}

It is indeed interesting to observe that frameless ALOHA exhibits a more complex behavior when performance are analyzed
versus \maxFrame. 
\rev{To explore this aspect, we report in Fig.~\ref{fig:contour}
the throughput and average \ac{AoI} pairs that can be achieved by tuning the maximum \ac{CP} duration. Specifically, we consider different values of \maxFrame, pick, for each setting, the optimal access probability $\pTx^*$, and plot $\tru^*(\maxFrame)$ and $\AoI^*(\maxFrame)$.}
The three different  curves in the figure refer to three different packet generation probabilities $\pAct\nodes\in\{0.6,0.8,1.0\}$, with $\nodes=200$. The lines illustrated our analytical results, whereas the markers correspond to simulation outcomes. \rev{The
\maxFrame\ intervals studied for each case are available in the figure caption, whereas the results obtained for three reference values, $\maxFrame=30$, $\maxFrame=100$, $\maxFrame=180$ are highlighted by special markers.}
Moreover, the maximum \ac{CP} duration leading to the minimum possible \ac{AoI} for each traffic generation level is explicitly reported in the plot for convenience.

Consider first the case $\pAct\nodes=0.8$, identified by the red curve (square markers) in Fig.~\ref{fig:contour}, and focus on throughput performance. For low values of \maxFrame, the system operates in the left region of the plot, providing low throughput. In this case, too short \acp{CP} hinder packet decoding, not allowing enough slots for \ac{SIC} to be fully efficient. By increasing
the maximum duration of the \ac{CP},  we can improve the throughput and approach
the elbow exhibited by the curve. 
After a certain point, though, a further increase of \maxFrame\ allows for the
decoding of only a limited additional number of users, and such diminishing-return behavior leads to a decrease in throughput. Notably, while a similar trend emerges also for the average \ac{AoI}, the impact of operating over excessively long \acp{CP} is far more pronounced. The rationale behind this lies in the dependency of \avgAoI\ on the inter-update time, i.e., the number of \acp{CP} between two updates as well as their duration in slots. From this standpoint, higher values of \maxFrame\ may reduce the former (increasing throughput), yet entail a larger average cost in terms of elapsed
slots over a \ac{CP}. While initially the first factor prevails, and \avgAoI\
improves together with \tru, the impact of longer \acp{CP} quickly turns out to
be detrimental in terms of AoI.

The analysis reveals then a fundamental trade-off between information freshness
and throughput, and implies that, for a given traffic generation rate, operating
the system at maximum throughput entails a degradation of performance in terms of \ac{AoI}. 
Such a behavior departs significantly from the one observed in plain ALOHA \cite{Yates17:AoI_SA}, 
and pinpoints a characteristic behavior of modern random access schemes
employing \ac{SIC}, first observed in \cite{Munari21_TCOM} for irregular repetition slotted ALOHA.

Similar results can be observed for $\pAct\nodes=0.6$ and $\pAct\nodes=1.0$, with the discussed effect becoming more pronounced for higher traffic.
Furthermore, longer \acp{CP} are needed to achieve better performance when
$\pAct\nodes$ increases, as the \ac{SIC} process benefits from additional slots
under harsher channel contention.
Indeed, the maximum throughput is attained for $\maxFrame=60$ slots when
$\pAct\nodes=0.6$ but for $\maxFrame=130$ slots when $\pAct\nodes=1.0$. 
A trade-off between \avgAoI\ and \tru\ emerges also when one looks at the nodes activation probability. 
Indeed, while the highest throughput among the considered setups is attained for
$\pAct\nodes=1.0$, the best results in terms of information freshness is
achieved for the lower level of contention $\pAct\nodes=0.8$.

\rev{To conclude our discussion, we present a performance comparison among frameless ALOHA and two benchmarks: a baseline slotted ALOHA scheme, and irregular repetition slotted ALOHA (IRSA) \cite{Liva11:IRSA}.}
For slotted ALOHA, we assume that a node immediately sends a newly generated update, without retransmissions, resulting in a probability of accessing the channel at each slot of~$\pAct$.
For the collision channel model under study, this leads to the classical throughput expression~\cite{Abramson77:PacketBroadcasting}
\begin{align}
  \tru_{\mathsf{sa}} = \pAct\nodes (1-\pAct)^{\nodes-1}.
  \label{eq:truSA}
\end{align}
Similarly, the average \ac{AoI} can be expressed in closed form as \cite{Munari21_TCOM,Yates17:AoI_SA}
\begin{align}
  \avgAoI_{\mathsf{sa}} = \frac{1}{2} + \frac{\nodes}{\tru_{\mathsf{sa}}}.
  \label{eq:ageSA}
\end{align}
Note that both metrics are \rev{optimized, i.e., maximum throughput, minimum AoI,} when the system operates at a channel load of $1$ packet per slot, i.e., corresponding to $\pAct\nodes=1$. 

\rev{On the other hand, IRSA is a well-known benchmark for modern random access schemes \cite{Berioli16_Now}, and operates similar to frameless ALOHA. Indeed, both protocols rely on having users transmit multiple copies of a packet and on the use of \ac{SIC} to resolve collisions. The main difference lies in the fact that a contention period in frameless ALOHA can be terminated by the receiver as soon as all users are decoded, whereas with IRSA the predefined end of the frame will always be reached. We refer the interested reader to \cite{Liva11:IRSA} for a detailed description of the operation procedures. Thanks to its ability to adaptively terminate the contention, frameless ALOHA is capable to improve throughput performance over IRSA \cite{Stefanovic12:Frameless}.\footnote{We note that these benefits come at the cost of a more frequent feedback, distributed by the receiver after each contention period. In this respect, IRSA can be less demanding, as sporadic beacons sent to the users can be sufficient to maintain frame alignment over time.} In turn, this feature can intuitively be beneficial also in terms of age of information, reducing the time between successive updates that a node can perform. To explore this aspect, we lean on the average IRSA AoI derivation in \cite{Munari21_TCOM}. To grant a fair comparison with the setup under study,  we assume that a transmitted packet has time stamp set to the start instant of the frame, and obtain
\begin{equation}
    \avgAoI_{\mathsf{irsa}} = \frac{d}{2} + \frac{\nodes}{\tru}
    \label{eq:age_irsa}
\end{equation}
where $d$ is the number of slots composing a frame and $\tru_{\mathsf{irsa}}$ is the protocol throughput. The result in \eqref{eq:age_irsa} highlights the important role played by the frame duration in IRSA. On the one hand, operating the protocol with longer frames can lead to better throughput performance \cite{Liva11:IRSA,Munari21_TCOM}, lowering the second component of $\Delta_{\mathrm{irsa}}$ in \eqref{eq:age_irsa}. On the other hand, larger values of $d$ contribute linearly to the growth of the first addend.}

\begin{figure}
\centering
\includegraphics[width=.65\textwidth]{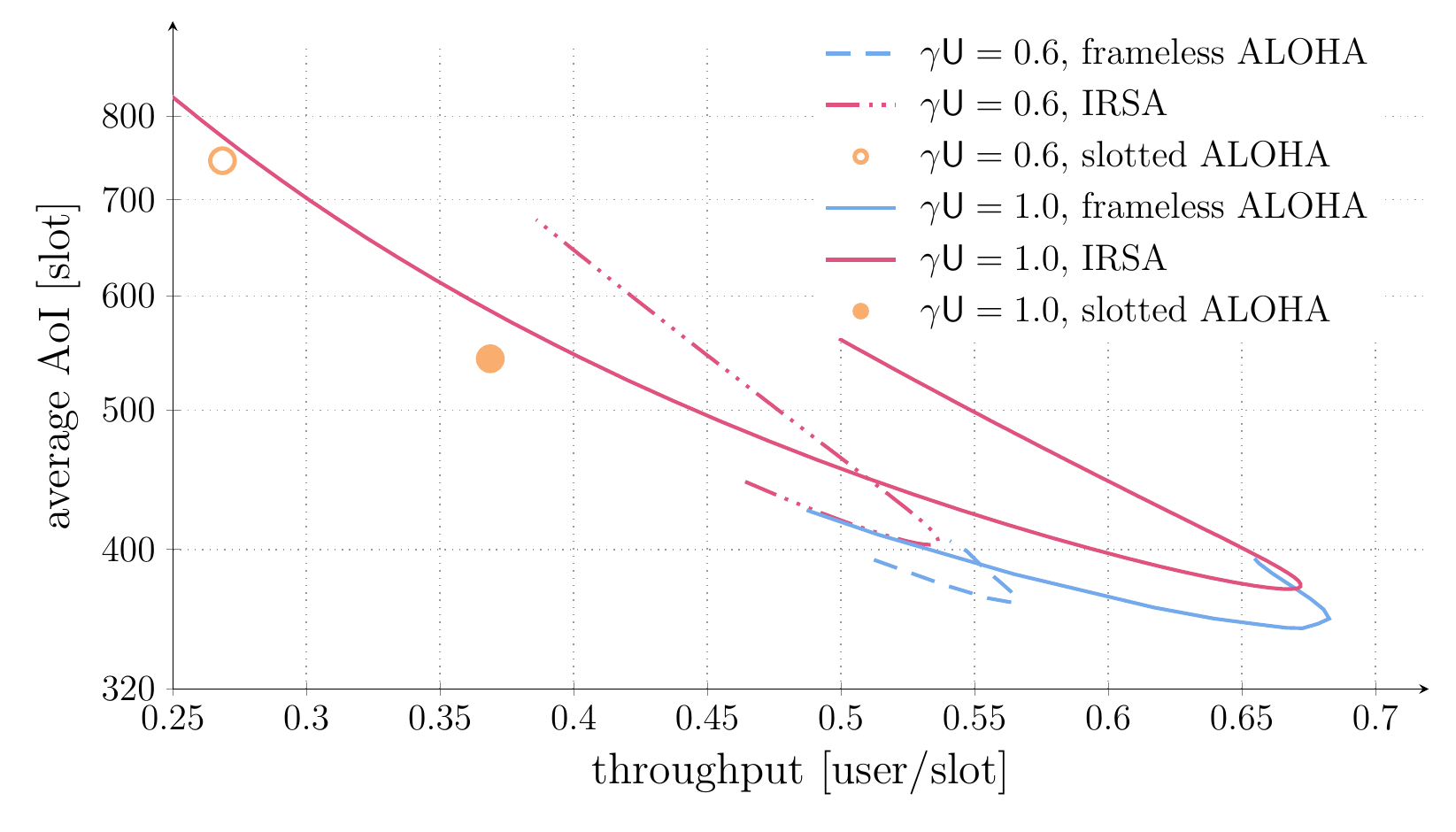}
\caption{Average AoI vs aggregate throughput for IRSA (magenta lines), frameless ALOHA (blue lines) and slotted ALOHA (circle markers) for two traffic generation intensities ($\pActGen\nodes=0.6$ and $\pActGen\nodes=1.0$). In all cases, $\nodes=200$ users were considered. For IRSA, a replica distribution $\Lambda(x) = 0.86x^3 + 0.14 x^8$ \cite{Graell18_Waterfall} was used, and each point in the curves denotes performance attained for a different frame duration, ranging between $30$ and $350$ slots. For frameless ALOHA, $\tru^*(\maxFrame)$ and $\AoI^*(\maxFrame)$ are shown, varying the maximum frame duration $\maxFrame$.}
\label{fig:irsa_comparison}
\end{figure}

\rev{The performance of the three schemes is compared in  Fig.~\ref{fig:irsa_comparison}, which reports the average AoI against the throughput for two different values of $\pActGen$, i.e. $\pActGen\nodes=0.6$ and $\pActGen\nodes=1.0$. For frameless ALOHA, results were obtained by varying the maximum contention period duration \maxFrame, and the points report $\AoI^*(\maxFrame)$ and $\tru^*(\maxFrame)$, akin to what done in Fig.~\ref{fig:contour}. As to IRSA, the protocol was operated using a replica distribution $\Lambda(x) = 0.86 x^3 + 0.14 x^8$ \cite{Graell18_Waterfall}, i.e., each node attempting transmission over a frame sends $3$ copies of its packet over the $d$ slots with probability $0.86$, whereas $8$ copies are sent with probability $0.14$. The curves show the performance for values of $d$ ranging from $30$ to $350$ slots.  Finally, the performance of slotted ALOHA is denoted by an empty ($\pActGen\nodes=0.6$) or filled ($\pActGen\nodes=1.0$) marker, whose coordiates are computed via~\eqref{eq:truSA} and~\eqref{eq:ageSA}.}

\rev{As expected, Fig.~\ref{fig:irsa_comparison} highlights that both advanced schemes clearly outperform the basic random access solution. More interestingly, frameless ALOHA improves over IRSA in both traffic configurations. Specifically, let us consider the minimum average AoI that the schemes can obtain when varying the contention duration. The plot shows that a metric reduction of $\sim 10\%$ and $\sim 7\%$ is achieved for $\pActGen\nodes = 0.6$ and $\pActGen\nodes = 1.0$, respectively. Notably, when the protocols are operated in these configurations, they offer very similar throughput. In other words, frameless ALOHA is capable of improving AoI without undergoing a penalty in terms of throughput when compared to IRSA. Along a similar line, one may want to operate the schemes so that throughput is maximized. In this situation, frameless ALOHA offers a minor throughput improvement of $\sim5\%$ and $\sim 2\%$ for $\pActGen\nodes = 0.6$ and $\pActGen\nodes = 1.0$, respectively. However, the average AoI in these settings is reduced by $\sim 9\%$ and $\sim 5\%$. Also in such operating conditions, therefore, resorting to frameless ALOHA can be beneficial.}

\rev{The outcomes of Fig.~\ref{fig:irsa_comparison} are complemented by Table~\ref{tab:saVsFrameless}, pinpointing the optimal results that can be obtained by the three considered protocols for four different values of $\pActGen\nodes$. The values reported for frameless ALOHA correspond to the maximum attainable throughput $\maxTru$ and minimum attainable average AoI $\minAoI$ that can be achieved by tuning \maxFrame.  Specifically, the results were obtained by optimizing over both the maximum \ac{CP} duration and the transmission probability \pTx, leading to $\maxTru = \max\nolimits_{\maxFrame} \tru^*(\maxFrame)$ and $\minAoI = \min\nolimits_{\maxFrame} \AoI^*(\maxFrame)$. The value of \maxFrame\ under which the shown performances are achieved (in general different for throughput and AoI) is also noted for completeness. Similarly, for IRSA, we report the optimal values for AoI and throughput obtained among all the considered frame durations.}
The improvements triggered by frameless ALOHA are apparent from the table, with an  average \ac{AoI} almost halved ($\sim 60\%$) compared to that of slotted ALOHA, and throughput gains up to $90\%$. We remark that such results are obtained simply by optimizing over the pair $(\maxFrame,\pTx)$. From this standpoint, the frameless approach offers additional room for improvements, which may pave the road for further gains. As an example, we provide next some initial considerations on how a dynamic adaptation of the maximum \ac{CP} duration may be leveraged to improve both AoI and throughput.

\begin{table*}
\centering
\caption{Performance comparison of slotted ALOHA, irregular repetition slotted ALOHA (IRSA) \cite{Liva11:IRSA} and frameless ALOHA, in terms of the maximum achievable throughput $\maxTru$ and the minimum achievable average AoI $\minAoI$ for different values of $\pAct\nodes$. For frameless ALOHA (IRSA), the maximum CP duration (frame duration) attaining the reported results is also given in parentheses. In all cases, $\nodes=200$.}
\label{tab:saVsFrameless}
{\renewcommand{\arraystretch}{1.2}
\begin{tabular}{c|c|c|c|c|c|c|c|c}
    \multicolumn{1}{c}{}&\multicolumn{2}{c}{$\pAct\nodes = 0.4$} & \multicolumn{2}{c}{$\pAct\nodes = 0.6$} & \multicolumn{2}{c}{$\pAct\nodes = 0.8$} & \multicolumn{2}{c}{$\pAct\nodes = 1.0$}\\[.4em]
     & $\maxTru$ &  $\minAoI$ & \multicolumn{1}{c}{$\maxTru$} &  $\minAoI$   & \multicolumn{1}{c}{$\maxTru$} &  $\minAoI$ & \multicolumn{1}{c}{$\maxTru$} &  $\minAoI$\\
     & [pkt/slot] &  [slot] & [pkt/slot] & [slot] & [pkt/slot] &  [slot] & [pkt/slot] &  [slot]\\
     \hline\hline
     slotted ALOHA & 0.2686 & 745.22 & 0.3300 & 606.59 & 0.3603 & 555.55 & 0.3688 & 542.79 \\
     \hline
     IRSA & 0.3838 & 537.65 & 0.5364 & 403.0279 & 0.6278 & 372.3377 & 0.6721 & 375.4708\\[-.4em]
     ($d$) & (35) & (31) & (65) & (56) & (113) & (103) & (160) & (151) \\
     \hline
     frameless ALOHA & 0.3987 & 503.54 & 0.5657 & 367.46 & 0.6399 & 351.67 & 0.6827 & 352.67\\[-.4em]
     (\maxFrame) & (30) & (30) & (60) & (45) & (100) & (70) & (130) & (110) \\
\end{tabular}
}
\end{table*}

\subsection{Performance Improvements via Early Termination of Contention Period}


%
It was shown in~\cite{Stefanovic13:RatelessAloha} that terminating the \ac{CP} after having decoded only a fraction of the contending users can be beneficial, at least in terms of  throughput.
The purpose of this section is to investigate whether introducing such a
termination criterion is also beneficial in terms of average \ac{AoI}.
In particular, in this section we compare two different contention termination strategies. 
The first one (baseline) is the one we considered so far, i.e., the \ac{CP} is
terminated after all users are decoded or after $\maxFrame$ slots have elapsed
since the beginning of the~\ac{CP}.
The second strategy, inspired by~\cite{Stefanovic13:RatelessAloha}, entails
terminating the \ac{CP} whenever the fraction of decoded users reaches $0.85$ or
after a total of $\maxFrame$ slots have elapsed. 
We refer to this strategy as \emph{early termination}.
Note that this strategy requires the receiver to know the number of contending
users. 
As argued in~\cite{Stefanovic13_ICC}, the number of contending users can be
estimated from the number of idle, singleton, and collided slots.

\begin{figure}
  \centering
  \includegraphics[width=0.65\columnwidth]{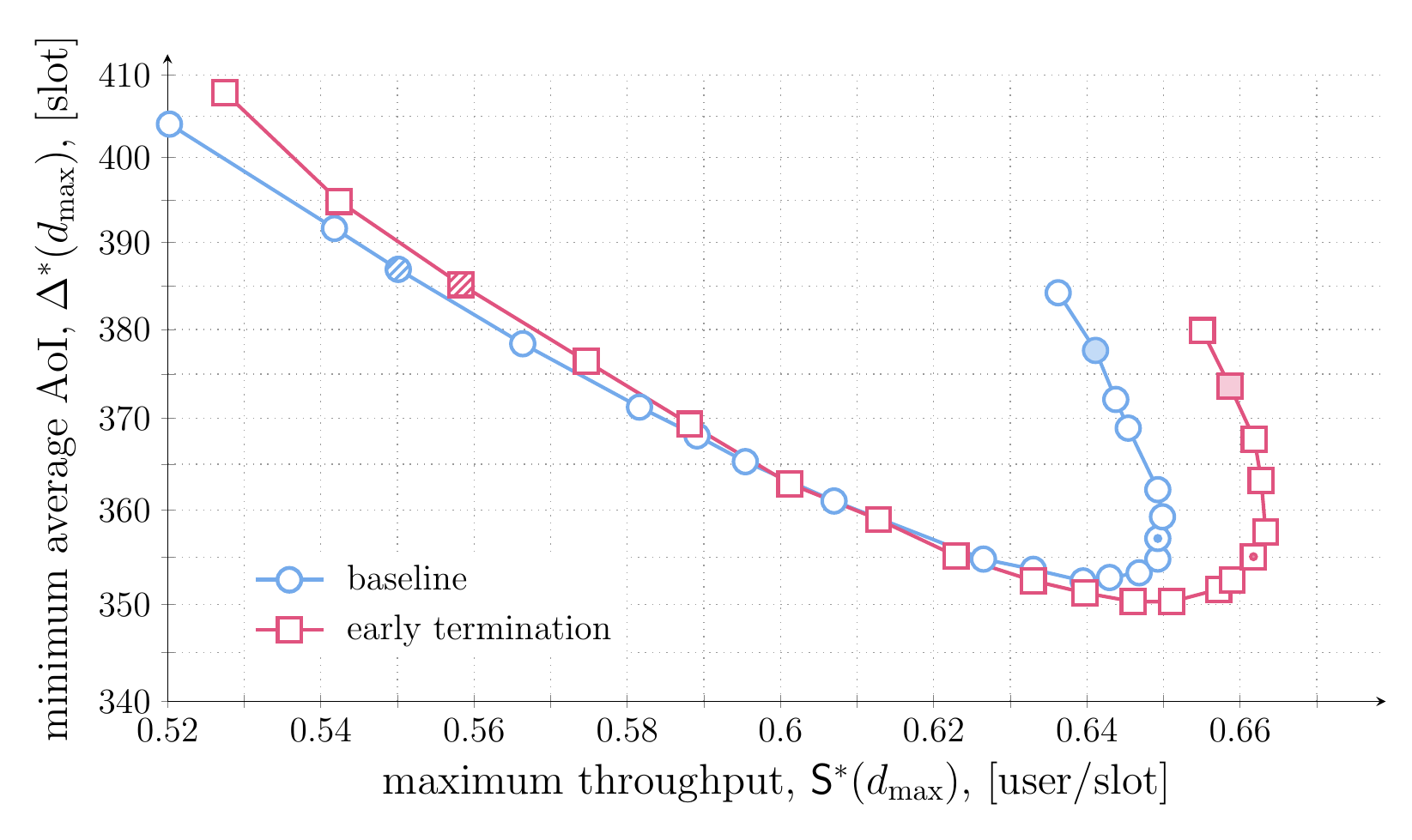}
  \caption{Maximum throughput and minimum average \ac{AoI} \rev{obtained when varying the maximum CP duration \maxFrame}, for $\pAct\nodes = 0.8$ and $\nodes=200$. \rev{Circle markers denote the baseline strategy where the contention is terminated after all contending users are decoded or after the maximum contention duration $\maxFrame$ is reached. In turn, the square markers  markers represent an early-termination strategy inspired 
  by~\cite{Stefanovic13:RatelessAloha}, in which the \ac{CP}
is terminated after at least $85\%$  of the contending users are decoded or after $\maxFrame$ slots have elapsed. For reference, striped markers indicate performance for $\maxFrame = 30$, markers filled with a point the case $\maxFrame = 100$, and color-filled markers denote $\maxFrame = 180$.}
}
  \label{fig:contour_dynamic}
\end{figure}

\rev{In Fig.~\ref{fig:contour_dynamic}, we show the maximum throughput $\tru^*(\maxFrame)$ and minimum
average AoI $\AoI^*(\maxFrame)$} for a setup with $\nodes=200$ users and  $\pAct\nodes = 0.8$ for the
two \ac{CP} termination strategies, and for values of $\maxFrame$ between $20$ and $220$ slots.
Note that, for small values of $\maxFrame$ (upper left region of the plot),
terminating the \ac{CP} before
decoding all users is detrimental both in terms of throughput and average
\ac{AoI}. On the contrary, for larger values of $\maxFrame$, this strategy is
beneficial for both performance metrics. 
\rev{The intuition behind these results is the following. Consider a contention period in which a fraction $f$ of the contending users is still unresolved. Let us consider two different groups of users: the first group contains the unresolved contending users, whereas the second group contains the pending users, i.e., the users who have an update to send in the next contention period. In terms of AoI, for the first group (unresolved users) it is beneficial to continue the contention so that these users have a chance of delivering a successful update. In contrast, for the second group (pending users) it is beneficial in terms of AoI to terminate the current contention so that they can immediately start contending to deliver their update. 
For small contention periods, it is better not to apply early termination, since the benefit of giving unresolved users the chance of successfully delivering an update outweighs the penalty incurred by having the pending users wait. 
In contrast, for large values of \maxFrame, once a substantial fraction of the users is resolved (e.g., $0.85$), it is better to terminate early the contention, so that pending users can immediately start contending to send their updates (although we condemn unresolved users to wait until their next update to reset their age). }

\section{Conclusions}
\label{sec:conclusions}

We studied the dynamic behavior of frameless ALOHA, focusing on the throughput
and the age of information (AoI) performance. The analysis is based on a
finite-length analysis of the successive interference cancellation process of
frameless ALOHA, and on a Markovian analysis of the system state evolution. We
characterized the
stability of the protocol via a drift analysis, which allowed us to
determine the presence of stable and unstable equilibrium points.
Finally, we provided an exact characterization of the AoI performance,
based on which we unveiled the impact of some key protocol parameters such as the maximum length
of the contention period, on the average AoI. Our results indicate that configurations of
parameters that maximize the throughput may result in a degradation of the AoI
performance.

\bibliographystyle{IEEEtran}
\bibliography{IEEEabrv,aloha,biblio_AoI}

\end{document}